%% file: main.tex
\documentclass[sigconf,obeyspaces,10pt]{acmart}

\renewcommand\footnotetextcopyrightpermission[1]{}
\usepackage{booktabs} 
\usepackage{makecell}

\usepackage{amsmath}
\usepackage{verbatim}
\usepackage{algorithmicx} 
\usepackage[plain]{algorithm}

\usepackage{algpseudocode}
\usepackage{pgfplots}
\usepackage{pgfplotstable}
\usepackage[center]{subfigure}
\usepackage{url}
\usetikzlibrary{patterns,shapes,arrows,positioning,fit,decorations.pathreplacing}
\usepackage{xcolor}

\usepgfplotslibrary{groupplots}
\usepackage{style/pgf-pie}

\usepackage{verbatim}
\usepackage{pifont}

\usepackage{multirow}
\usepackage{enumitem}

\theoremstyle{definition}
\newtheorem{example}{Example}
\newtheorem{definition}{Definition}

\setcopyright{none} 

\acmDOI{12.345/123_4}

\acmISBN{123-4567-89-012/03/04}

\acmConference[SIGMOD'19]{ACM SIGMOD conference}{June 2019}{Amsterdam, The Netherlands} 

\begin{document}
\title{Verifying Text Summaries of Relational Data Sets}

\author{Saehan Jo\footnotemark[1], Immanuel Trummer\footnotemark[1], Weicheng Yu\footnotemark[1], Daniel Liu\footnotemark[1],\\Xuezhi Wang\footnotemark[2], Cong Yu\footnotemark[2], Niyati Mehta\footnotemark[1]}
\affiliation{
	\institution{Cornell University\footnotemark[1], Google Research\footnotemark[2]}
}
\email{{sj683, itrummer, wy248, dl596, nbm44}@cornell.edu, {xuezhiw, congyu}@google.com}

\begin{abstract}
We present a novel natural language query interface, the AggChecker, aimed at text summaries of relational data sets. The tool focuses on natural language claims that translate into an SQL query and a claimed query result. Similar in spirit to a spell checker, the AggChecker marks up text passages that seem to be inconsistent with the actual data. At the heart of the system is a probabilistic model that reasons about the input document in a holistic fashion. Based on claim keywords and the document structure, it maps each text claim to a probability distribution over associated query translations. By efficiently executing tens to hundreds of thousands of candidate translations for a typical input document, the system maps text claims to correctness probabilities. This process becomes practical via a specialized processing backend, avoiding redundant work via query merging and result caching. Verification is an interactive process in which users are shown tentative results, enabling them to take corrective actions if necessary. 

Our system was tested on a set of 53 public articles containing 392 claims. Our test cases include articles from major newspapers, summaries of survey results, and Wikipedia articles. Our tool revealed erroneous claims in roughly a third of test cases. A detailed user study shows that users using our tool are in average six times faster at checking text summaries, compared to generic SQL interfaces. In fully automated verification, our tool achieves significantly higher recall and precision than baselines from the areas of natural language query interfaces and fact-checking.
\end{abstract}

\maketitle


\section{Introduction}
\label{introSec}
\input{sections/intro2.tex}

\section{Problem Statement}
\label{problemSec}
\input{sections/problem.tex}

\section{System Overview}
\label{overviewSec}
\input{sections/overview.tex}


\section{Keyword Matching}
\label{matchingSec}
\input{sections/matching.tex}

\section{Probabilistic Model}
\label{modelSec}
\input{sections/model.tex}

\section{Query Evaluation}
\label{querySec}
\input{sections/query.tex}

\section{Experimental Evaluation}
\label{experimentsSec}
\input{sections/experiments2.tex}

\section{Related Work}
\label{relatedSec}
\input{sections/related.tex}

\section{Conclusion}
\label{conclusionSec}
\input{sections/conclusion.tex}


\bibliographystyle{ACM-Reference-Format}
\bibliography{sample-bibliography} 


\appendix

\input{sections/appendix3.tex}

\end{document}

%% file: sections/intro2.tex


Relational data is often summarized by text. Examples range from newspaper articles by data journalists to experimental papers summarizing tabular results. We focus on the novel problem of verifying, in an automated fashion, whether text claims are consistent with the actual database. Corresponding tools can help text authors (e.g., a business analysts summarizing a relational data set with sales) to polish their text before publication. Alternatively, we enable users to verify text written by third parties (e.g., a lector could check an article before publication). Our analysis of hundreds of real-world claims on relational data, discussed in detail later, shows the potential and need for such tools.

We present a tool for verifying text summaries of relational data sets. Our tool resembles a spell checker and marks up claims that are believed to be erroneous. We focus on natural language claims that can be translated into an SQL query and a claimed query result. More precisely, we focus on claims that are translated into aggregation queries on data subsets. Hence the name of our system: AggChecker. Our analysis shows that this claim type is at the same time very common and error-prone. The following example illustrates the concept.

\begin{example}
	\label{claimExa}
	Consider the passage \textit{``There were only four previous lifetime bans in my database - three were for repeated substance abuse''} taken from a 538 newspaper article~\cite{53814-2}. It contains two claims that translate into the SQL queries {\scshape SELECT Count(*) FROM nflsuspensions WHERE Games = `indef'} (with claimed result `four')
	and {\scshape SELECT Count(*) FROM nflsuspensions WHERE Games = `indef' AND Category = `substance abuse, repeated offense'} (with claimed result `three') on the associated data set. Our goal is to automatically translate text to queries, to evaluate those queries, and to compare the evaluation result against the claimed one.
\end{example}

Internally, the system executes the following, simplified process to verify a claim. First, it tries to translate the natural language claim into an SQL query reflecting its semantics. Second, it executes the corresponding query on the relational database. Third, it compares the query result against the value claimed in text. If the query result rounds to the text value then the claim has been verified. Otherwise, the claim is considered erroneous. Color markup indicates the verification result to users. Additionally, users may obtain information on the verification process and can take corrective actions if necessary (similar to how users correct erroneous spell checker markup).

The most challenging step is of course the translation of a natural language claim into an SQL query. There has been a recent surge in research on natural language query interfaces~\cite{Arbor2016, Li2014, Saha2016}. Our work is situated in that space. We address however a novel problem variant that has not been treated before. Table~\ref{ctVsqt} summarizes differences between natural language querying (NLQ) and natural language claim to query translation (CTQ). In case of NLQ, the input is a single natural language query. In our case, we use as input an entire text document containing multiple, related claims. Typical NLQ approaches output a single result query (whose result is presented to the user). In our case, we map each claim to a probability distribution over corresponding SQL queries. As discussed in more detail later, we can calculate correctness probabilities for a given claim based on the latter distribution.

Our goal was to create a system that works ``out of the box'' for a new input text and database. We do not assume that training samples are available for a given database. In that, our intent is similar to other recent work in the area of natural language query interfaces~\cite{Arbor2016}. We also do not require users to annotate data by hand to enable claim to query translation.

Those restrictions make it very hard to translate claims to queries. Among the challenges we encountered when studying real-world test cases are the following. First, the claim sentence itself is often missing required context. This context can only be found when analyzing preceding paragraphs or headlines (assuming a hierarchical input text document). Second, claim sentences often contain multiple claims which make it hard to associate sentence parts to claims. Third, claim sentences are often long and contain parts which do not immediately correspond to elements in the associated query. This makes it hard to map the claim sentence parse tree to a similar SQL query tree. Fourth, data sets often contain entries (e.g., abbreviations) that are not found immediately in the claim text. Altogether, this makes it hard to map claim sentences unambiguously to database elements. Our system exploits the following main ideas to cope with those challenges.

\begin{table}
\caption{Natural language querying versus claim to query translation.\label{ctVsqt}}
\begin{small}
\begin{tabular}{p{1.75cm}p{2.6cm}p{3cm}}
\toprule[1pt]
\textbf{Variant} & \textbf{Input} & \textbf{Output} \\
\midrule[1pt]
Natural & One natural & One SQL query \\
language & language query & \\
querying & & \\
\midrule
Translating & Hierarchical text & Probability distribution \\
claims to & containing multiple & over claim translations \\
queries & related claims & for each claim \\
\bottomrule[1pt]
\end{tabular}
\end{small}
\end{table}

\textbf{Learn to Live with Uncertainty.} Typical natural language query interfaces need to produce a single query. For us, it is sufficient to know that none of the likely query candidates evaluates to a claimed result value. Hence, our system still works even if we cannot precisely translate specific text claims.

\textbf{Exploit Semantic Correlations.} Claims in the same input text are often semantically correlated. Hence, they translate into queries with similar characteristics. We exploit that fact by learning, via an iterative expectation-maximization approach, a document-specific prior distribution over queries (also called ``topic'' in the following). We integrate priors when selecting likely query candidates as translations.

\textbf{Exploit Claim Results.} We are given a query description as well as a query result. In practice, accurate claims are more likely than inaccurate claims. Hence, we consider a match between query result and claimed result a strong (but not decisive!) signal, increasing the likelihood of that query candidate. As we share knowledge between different claims (see the previous point), a clear signal received for one claim ideally resolves ambiguities for many others.

\textbf{Massive-Scale Query Evaluations.} To leverage those signals, we must evaluate candidate queries at a massive scale. Candidates are formed by combining relevant query fragments (such as aggregation functions, columns, and predicates). Query fragments are generated via a keyword-based high-recall heuristic by comparing database entries and claim keywords. We routinely evaluate ten thousands of query candidates per claim. To make this approach practical, we use an execution engine that merges execution of similar queries (among several other techniques) to increase efficiency.

If all else fails, we rely on the user to take corrective actions (i.e., the system is used in semi-automated instead of fully automated verification mode). The benefit of using the system is then similar to the benefit of using a spell checker. AggChecker resolves most cases and allows users to focus on difficult cases, thereby saving time in verification. We evaluated our system on a variety of real-world test cases, containing 392 claims on relational data sets. Our test cases cover diverse topics and derive from various sources, reaching from Wikipedia to New York Times articles. We generated ground truth claim translations by hand and contacted the article authors in case of ambiguities. We identified a non-negligible number of erroneous claims, many of which are detected by our system. We compare against baseline systems and perform a user study. The user study demonstrates that users verify documents significantly faster via AggChecker than via standard query interfaces. 

In summary, our original scientific contributions are the following:

\begin{itemize}
\item We introduce the problem of translating natural language claims on relational data to SQL queries, without using prior training or manual annotations.
\item We propose a first corresponding system whose design is tailored to the particularities of our scenario.
\item We compare our system against baselines in fully automated checking as well as in a user study.
\end{itemize}

The remainder of this paper is organized as follows. We formalize our problem model in Section~\ref{problemSec}. Next, we give an overview of our system in Section~\ref{overviewSec}. The following three sections describe specific components of our system: keyword matching, probabilistic reasoning, and massive-scale candidate query evaluations. After that, we present experimental results in Section~\ref{experimentsSec}. Finally, we compare against related work in fact-checking and natural language query interfaces in Section~\ref{relatedSec}. In the appendix, we provide more experimental results and list all our test cases.

%% file: sections/problem.tex
We introduce our problem model and related terminology. We generally assume a scenario where we have a relational \textit{Database} together with a natural language \textit{Text} summarizing it. The relational database might be optionally associated with a data dictionary (mapping database elements such as columns and values to text descriptions). The text document may be semi-structure, i.e. it is organized as a hierarchy of sections and subsections with associated headlines. Also, the text contains \textit{Claims} about the database. 
\begin{definition}A \textbf{Claim} is a word sequence from the input text stating that evaluating a query $q$ on the associated database $D$ yields a rounded result $e$. We focus on SQL queries with numerical results ($e\in\mathbb{R}$). We call $q$ also the \textit{Matching Query} or \textit{Ground Truth Query} with regards to the claim. A claim may be a sentence part or a sentence (one sentence may contain multiple claims). A claim is \textit{Correct} if there is an admissible rounding function $\rho:\mathbb{R}\rightarrow\mathbb{R}$ such that the rounded query results equals the claimed value (i.e., $\rho(q(D))=e$). 
\end{definition}

We currently consider rounding to any number of significant digits as admissible. The approach presented in the next sections can be used with different rounding functions as well. We focus on \textit{Simple Aggregate Queries}, a class of claim queries defined as below.
\begin{definition}A \textbf{Simple Aggregate Query} is an SQL query of the form {\scshape SELECT Fct(Agg) FROM T1 E-JOIN T2 ... WHERE C1 = V1 AND C2=V2 AND ...}, calculating an aggregate over an equi-join between tables connected via primary key-foreign key constraints. The where clause is a conjunction of unary equality predicates. 
\end{definition}

Claims of this format are very popular in practice and at the same time error-prone (see Section~\ref{experimentsSec}). Currently, we support the following aggregation functions: {\scshape Count, Count Distinct, Sum, Average, Min, Max, Percentage,} and {\scshape Conditional Probability}\footnote{For conditional probability, we assume that the first predicate is the condition and the rest form the event. That is, ({\scshape SELECT ConditionalProbability(Agg) FROM T1 E-JOIN T2 ... WHERE C1 = V1 AND C2=V2 AND ...}) = ({\scshape SELECT Count(Agg) FROM T1 E-JOIN T2 ... WHERE C1 = V1 AND C2=V2 AND ...}) * 100 / ({\scshape SELECT Count(Agg) FROM T1 E-JOIN T2 ... WHERE C1 = V1}).} (we plan to gradually extend the scope). The ultimate goal would be to perform purely \textit{Automatic Aggregate-Checking} (i.e., given a text document and a database, identify claims automatically and decide for each one whether it is correct). This would however require near-perfect natural language understanding which is currently still out of reach. Hence, in this paper, we aim for \textit{Semi-Automatic Aggregate-Checking} in which we help users to verify claims without taking them out of the loop completely. 
\begin{definition}
Given input $\langle T,D\rangle$, a text $T$ and a database $D$, the goal of \textbf{Semi-Automatic Aggregate-Checking} is to identify claims and to map each claim $c$ to a probability distribution $Q_c$ over matching queries. This probability distribution can be exploited by a corresponding user interface to quickly verify text in interaction with the user. The quality of a corresponding approach can be measured based on how often the top-x likely query candidates in $Q_c$ contain the matching query.
\end{definition}

%% file: sections/overview.tex
\begin{figure}
	\centering
	\includegraphics[width=0.45\textwidth]{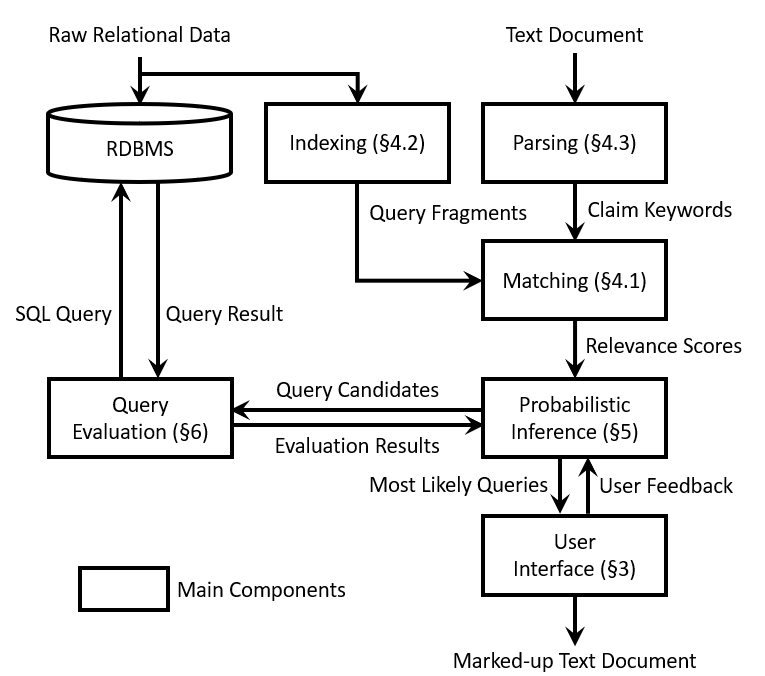}
	\caption{Overview of AggChecker system.\label{architectureFig}}
\end{figure}

Figure~\ref{architectureFig} shows an overview of the AggChecker system. The input to the AggChecker consists of two parts: a relational data set and a text document, optionally enriched with HTML markup highlighting the text structure. The text contains claims about the data. Our goal is to translate natural language claims into pairs of SQL queries and claimed query results. The process is semi-automated and relies on user feedback to resolve ambiguities. Finally, we enrich the input text with visual markup, identifying claims that are inconsistent with the data.

\begin{figure}
	\centering
	\subfigure[Raw relational data.\label{exampleTableFig}]
	{\includegraphics[width=0.240\textwidth]{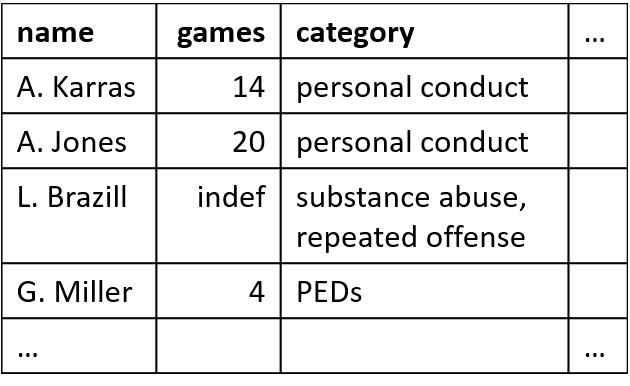}}
	\hspace{0.01\textwidth}
	\subfigure[Text document. Claims colored in blue. We focus on claimed result `one'.\label{exampleTextFig}]
	{\includegraphics[width=0.216\textwidth]{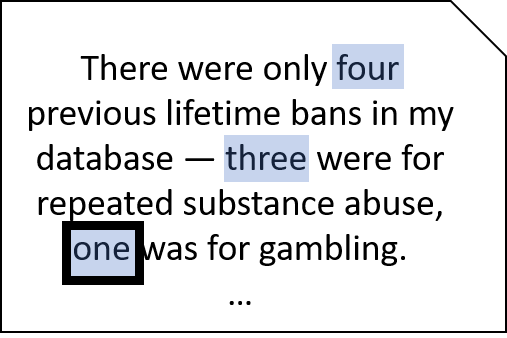}}
	
	\subfigure[Query fragments.\label{exampleFragmentsFig}]
	{\includegraphics[width=0.240\textwidth]{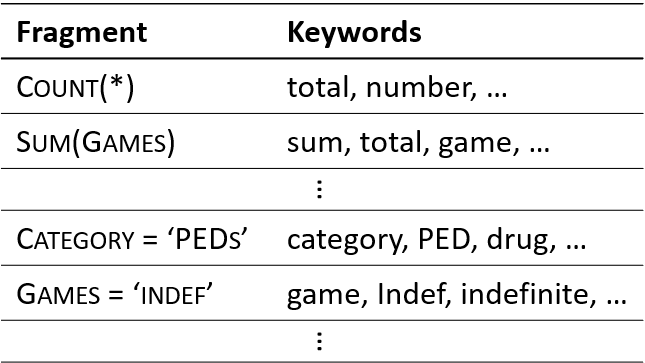}}
	\hspace{0.01\textwidth}
	\subfigure[Keywords and their weights (specific to claimed result `one').\label{exampleKeywordsFig}]
	{\includegraphics[width=0.216\textwidth]{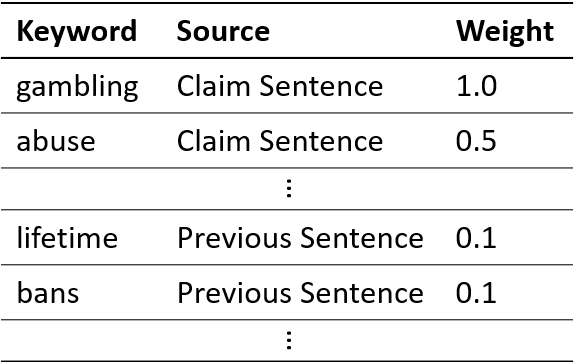}}
	
	\subfigure[Probability distribution and evaluation results of query candidates. Green indicates a match between query result and claimed result while red indicates the opposite (specific to claimed result `one').\label{exampleProbabilityFig}]
	{\includegraphics[width=0.48\textwidth]{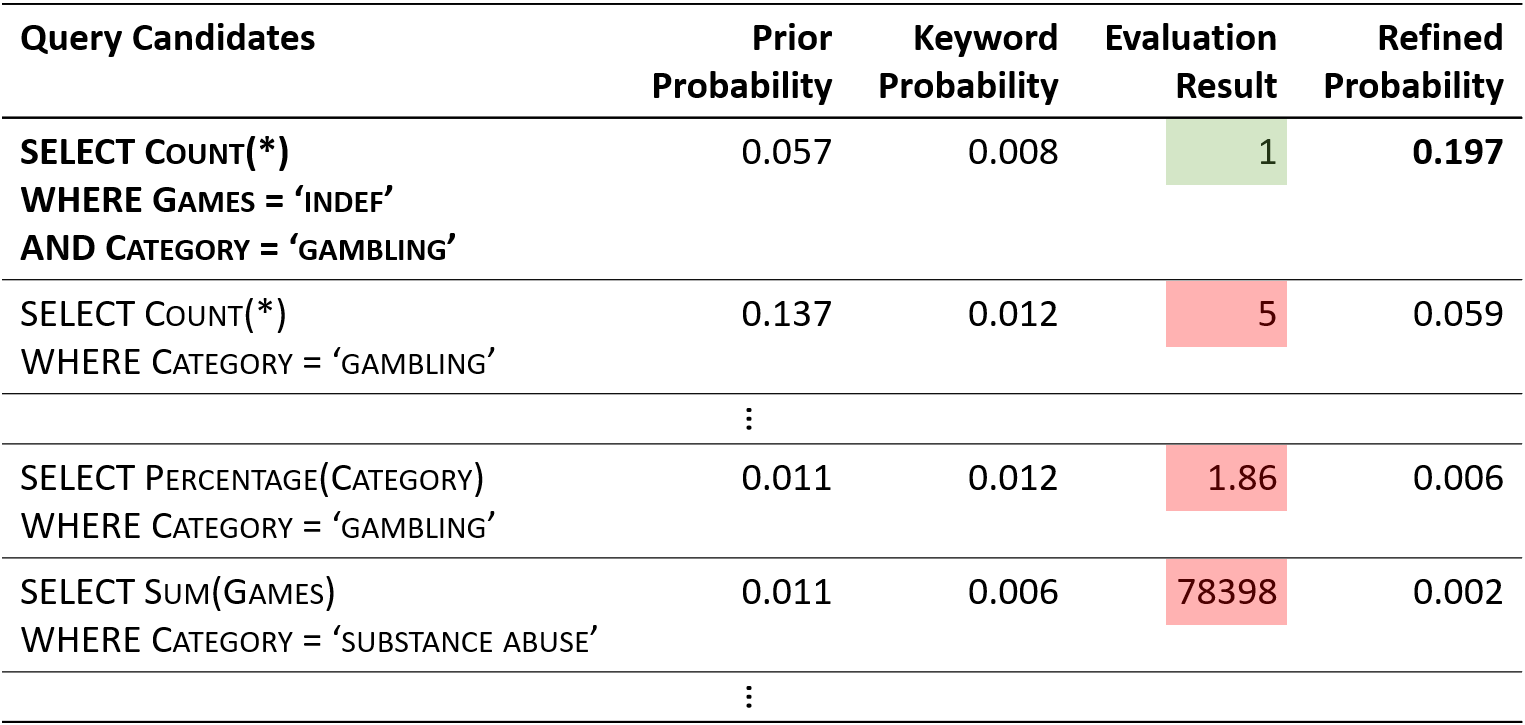}}
	\caption{Running example of AggChecker system.\label{exampleFig}}
\end{figure}

For each newly uploaded data set, we first identify relevant query fragments (see Figure~\ref{exampleFragmentsFig}). The system focuses on \textit{Simple Aggregate Queries} as defined in Section~\ref{problemSec}. Query fragments include aggregation functions, aggregation columns, or unary equality predicates that refer to columns and values in the data set. We associate each query fragment with keywords, using names of identifiers within the query fragment as well as related keywords that we identify using WordNet~\cite{Miller95, Fellbaum98}. We index query fragments and the associated keywords via an information retrieval engine (we currently use Apache Lucene~\cite{Lucene}). 


Next, we parse the input text using natural language analysis tools such as the Stanford parser~\cite{ManningS14}. We identify potentially check-worthy text passages via simple heuristics and rely on user feedback to prune spurious matches. Then, we associate each claim with a set of related keywords (see Figure~\ref{exampleKeywordsFig}). We use dependency parse trees as well as the document structure to weight those keywords according to their relevance. We query the information retrieval engine, indexing query fragments, using claim keywords as queries. Thereby we obtain a ranked set of query fragments for each claim.

Query fragments with relevance scores form one out of several inputs to a probabilistic model. This model maps each text claim to a probability distribution over SQL query candidates, representing our uncertainty about how to translate the claim (see Figure~\ref{exampleProbabilityFig}). The model considers the document structure and assumes that claims in the same document are linked by a common theme. The document theme is represented via model parameters capturing the prior probabilities of certain query properties. We infer document parameters and claim distributions in an iterative expectation-maximization approach. Furthermore, we try to resolve ambiguities in natural language understanding via massive-scale evaluations of query candidates. The AggChecker uses evaluation strategies such as query merging and caching to make this approach practical (we currently use Postgres~\cite{Postgres} to evaluate merged queries). We typically evaluate several tens of thousands of query candidates to verify one newspaper article.

\begin{example}
Figure~\ref{exampleFig} provides a concrete running example demonstrating the inputs and outputs of the main components. Figure~\ref{exampleTableFig} depicts the raw relational data where query fragments and their associated keywords are extracted as in Figure~\ref{exampleFragmentsFig}. Figure~\ref{exampleTextFig} illustrates a text passage from a 538 newspaper article~\cite{53814-2}. It contains three claimed results (colored in blue) where we focus on the claimed result `one' in this example. In Figure~\ref{exampleKeywordsFig}, we extract relevant keywords for this claimed result and weigh them based on the text structure. Then, we calculate relevance scores for pairs of query fragments and claims based on their keywords. The probabilistic model takes into account the relevance scores as well as two other inputs to infer the probability distribution over query candidates. Figure~\ref{exampleProbabilityFig} captures this concept. First, `Keyword Probability' is derived from the relevance scores. Second, `Prior Probability' encapsulates the model parameters that embrace all claims in the text document in a holistic fashion. Third, green or red color under `Evaluation Result' shows whether the query result matches the value claimed in text. Lastly, `Refined Probability' illustrates the final probability distribution over query candidates, considering all three inputs. After expectation-maximization iterations have converged, the system verifies the claim according to the query with the highest probability. We provide more detailed explanations in each section where all following examples refer to this figure.
\end{example}


\begin{figure}
\centering
\subfigure[Marked up claims after initial processing.\label{markupFig}]
{\includegraphics[width=0.19\textwidth]{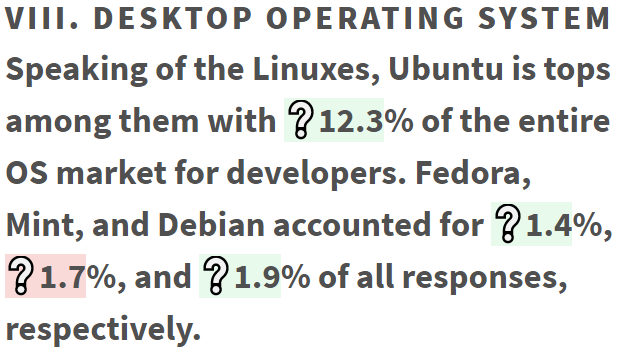}}
\subfigure[Query description shown upon hovering.\label{queryFig}]
{\includegraphics[width=0.19\textwidth]{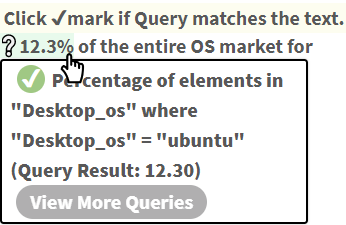}}
\subfigure[Select query from top-5 likely candidates.\label{topFig}]
{\includegraphics[width=0.19\textwidth]{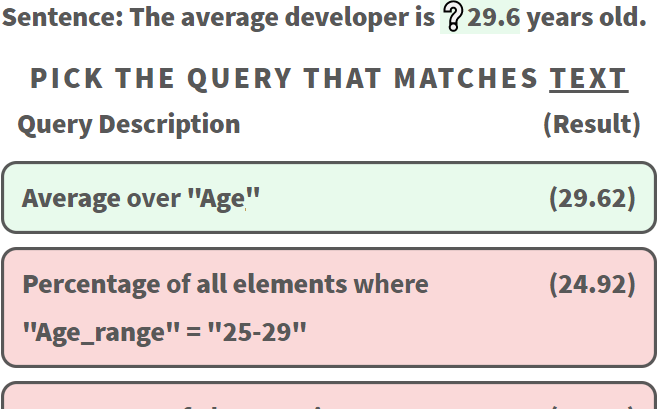}}
\subfigure[Construct query by selecting fragments.\label{fragmentsFig}]
{\includegraphics[width=0.19\textwidth]{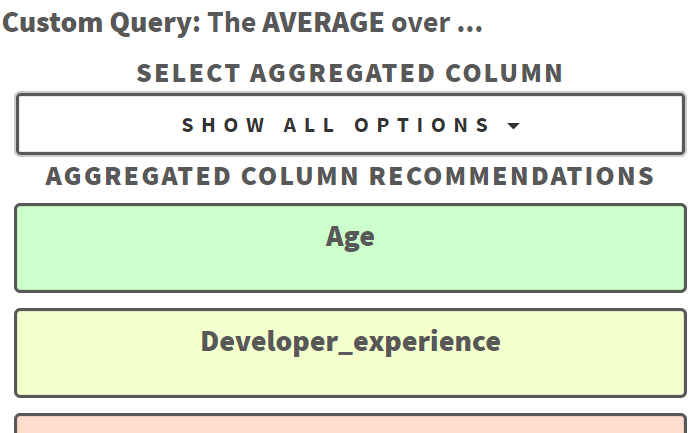}}
\caption{Screenshots from AggChecker interface.\label{interfaceFig}}
\end{figure}

After an automated verification stage, the system shows tentative verification results to the user. Claims are colored based on their probability of being erroneous (see Figure~\ref{markupFig}). Users can hover over a claim to see a natural language description of the most likely query translation (see Figure~\ref{queryFig}) and may correct the system if necessary. Alternatively, users may pick among the top-k most likely query candidates (see Figure~\ref{topFig}) or assemble the query from query fragments with high probability (see Figure~\ref{fragmentsFig}).

%% file: sections/matching.tex
In the first processing phase, we extract query fragments and claim keywords from the inputs and match them together to calculate relevance scores.

\subsection{Keyword Matching Overview}
We calculate relevance scores for pairs of claims and query fragments. The higher the relevance score, the more likely the fragment to be part of the query matching the claim. We consider aggregation functions, aggregation columns, and predicate parts as query fragments. Given an input database, we can infer all potentially relevant query fragments (i.e., we introduce an equality predicate fragment for each literal in the database, an aggregation column fragment for each column containing numerical values etc.). Furthermore, we can associate query fragments with relevant keywords (e.g., the name of a literal, as well as the name of the containing column and synonyms for a fragment representing an equality predicate). 

On the other side, we can associate each claim in the input text with relevant keywords, based on the document structure. Having query fragments and claims both associated with keyword sets, we can use methods from the area of information retrieval to calculate relevance scores for specific pairs of query fragments and claims. For instance, we use Apache Lucene in our current implementation, indexing keyword sets for query fragments and querying with claim-specific keyword sets. While keyword-based relevance scores are inherently imprecise, they will form one out of several input signals for the probabilistic model described in the next section. The latter model will associate each claim with a probability distribution over query candidates.



\begin{algorithm}[t]
\renewcommand{\algorithmiccomment}[1]{// #1}
\begin{small}
\begin{algorithmic}[1]
\State \Comment{Calculates claim-specific relevance scores for query}
\State \Comment{fragments on database $D$ for claims $C$ from text $T$.}
\Function{KeywordMatch}{$T,D,C$}
\State \Comment{Index query fragments by keywords}
\State $I\gets$\Call{IndexFragments}{$D$}
\State \Comment{Iterate over claims}
\For{$c\in C$}
\State \Comment{Retrieve claim-specific keyword context}
\State $K\gets$\Call{ClaimKeywords}{$c,T$}
\State \Comment{Calculate relevance scores for query fragments}
\State $S_c\gets$\Call{IrRetrieval}{$I,K$}
\EndFor
\State \Return{$\{S_c|c\in C\}$}
\EndFunction
\end{algorithmic}
\end{small}
\caption{Maps each claim to potentially relevant query fragments based on keywords.\label{matchingMainAlg}}
\end{algorithm}

\begin{algorithm}[t]
\renewcommand{\algorithmiccomment}[1]{// #1}
\begin{small}
\begin{algorithmic}[1]
	\State \Comment{Extract keywords for claim $c$ from text $T$.}
	\Function{ClaimKeywords}{$c,T$}
	\State \Comment{Initialize weighted keywords}
	\State $K\gets\emptyset$
	\State \Comment{Add keywords in same sentence}
	\For{$word\in c.sentence$}
	\State $weight\gets1/$\Call{TreeDistance}{$word,c$}
	\State $K\gets K\cup\{\langle word,weight\rangle\}$
	\EndFor
	\State \Comment{Add keywords of sentences in same paragraph}
	\State $m\gets\min\{1/$\Call{TreeDistance}{$k,c$}$|k\in c.sentence\}$
	\State $K\gets K\cup \{\langle k,0.4m\rangle|k\in c.prevSentence\}$
	\State $K\gets K\cup \{\langle k,0.4m\rangle|k\in c.paragraph.firstSentence\}$
	\State \Comment{Add keywords in preceding headlines}
	\State $s\gets c.containingSection$
	\While{$s\neq\mathbf{null}$}
	\State $K\gets K\cup \{\langle k,0.7m\rangle|k\in s.headline.words\}$
	\State $s\gets s.containingSection$
	\EndWhile
	\State \Return{$K$}
	\EndFunction
\end{algorithmic}
\end{small}
\caption{Extracts a set of keywords for a claim.\label{keywordAlg}}
\end{algorithm}

Algorithm~\ref{matchingMainAlg} summarizes the process by which relevance scores are calculated. It relies on sub-functions for indexing query fragments derived from the database (Function~\textproc{IndexFragments}) and for extracting keyword sets for claims (Function~\textproc{ClaimKeywords}). 

\subsection{Indexing Query Fragments}
When loading a new database, we first form all potentially relevant query fragments. Function~\textproc{IndexFragments} (we describe its implementation without providing pseudo-code) traverses the database in order to form query fragments that could be part of a claim query. We consider three types of query fragments: aggregation functions, aggregation columns, and equality predicates. All aggregation function specified in the SQL standard are potentially relevant (we could easily add domain-specific aggregation functions). We consider all numerical columns in any table of the database as aggregation columns (in addition, we consider the ``all column'' {\scshape *} as argument for count aggregates). Finally, we consider all equality predicates of the form $c=v$ where $c$ is a column and $v$ a value that appears in it.


We associate each query fragment with a set of relevant keywords. Keyword sets are indexed via an information retrieval engine (together with a reference to the corresponding fragment). We associate each standard SQL aggregation function with a fixed keyword set. The keywords for aggregation columns are derived from the column name and the name of its table. Column names are often concatenations of multiple words and abbreviations. We therefore decompose column names into all possible substrings and compare against a dictionary. Furthermore, we use WordNet to associate each keyword that appears in a column name with its synonyms. The keywords for an equality predicate of the form $c=v$ are derived from the column name $c$ (and from the name of the containing table) as well as from the name of value $v$. Finally, the AggChecker also offers a parser for common data dictionary formats. A data dictionary associates database columns with additional explanations. If a data dictionary is provided, we add for each column the data dictionary description to its associated keywords.

\subsection{Extracting Keywords from Text}
Next, we associate each claim in the input text with a weighted set of keywords. More precisely, we iterate over each number in the input text that is likely to represent a claimed query result. We describe in Section~\ref{overviewSec} how they are identified. Algorithm~\ref{keywordAlg} associates each such claim with weighted keywords, extracted from the containing text. First, we consider keywords in the claim sentence itself (i.e., the sentence in which the claimed result number is found). One sentence might contain multiple claims and we must decide what keywords are most relevant to one specific claim. For that, we construct a dependency parse tree of the claim sentence. We make the simplifying assumption that sentence parts are more closely related, the lower their distance (i.e., number of tree edges to traverse) is in the parse tree. Hence, for each numerical aggregate representing the result of a claim, we weight the surrounding keywords based on their distance from the numerical aggregate in the dependency tree (denoted by \textproc{TreeDistance} in Algorithm~\ref{keywordAlg}).

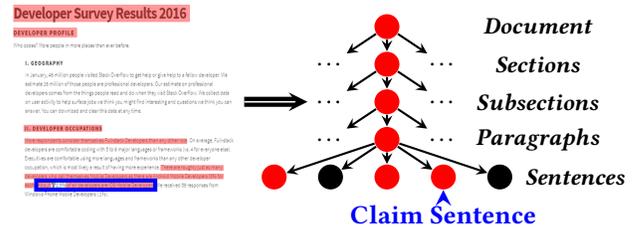
\begin{figure}
\begin{tikzpicture}[layer/.style={align=left, font=\bfseries\itshape\small}, treeNode/.style={fill=red, minimum width=0.25cm, circle}, child/.style={thick, -stealth}]

\node[layer] (text) at (-3.5,-1) {\includegraphics[scale=0.12]{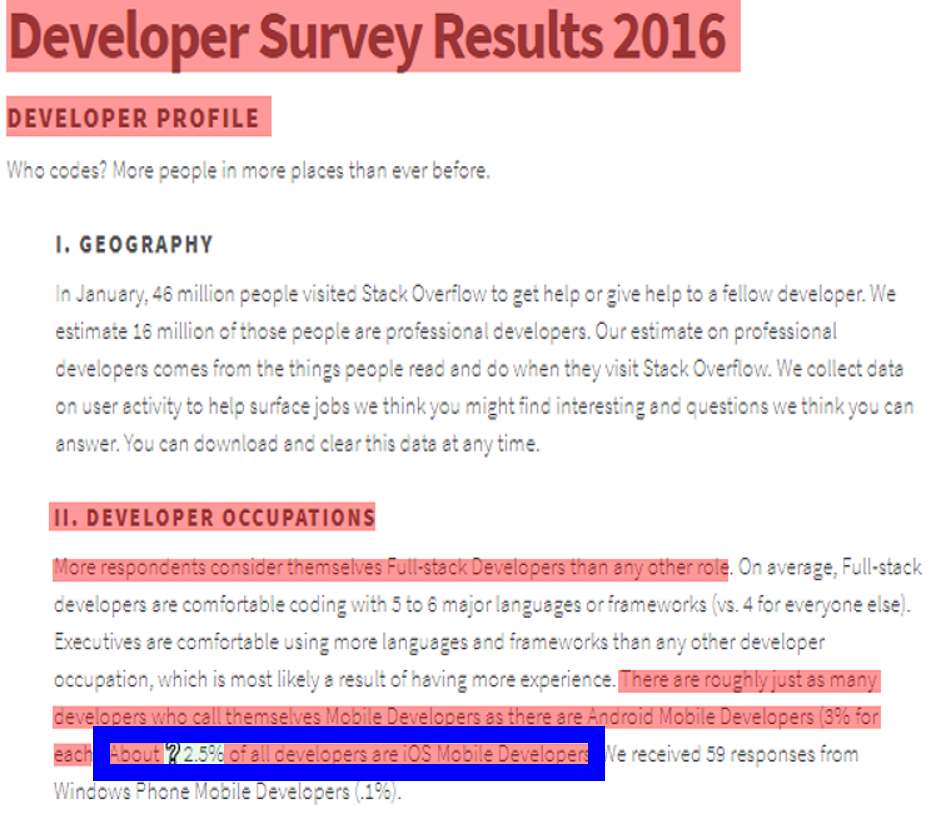}};

\node[layer] at (2,0) {Document};
\node[layer] at (2,-0.5) {Sections};
\node[layer] at (2,-1) {Subsections};
\node[layer] at (2,-1.5) {Paragraphs};
\node[layer] at (2.5,-2) {Sentences};

\node[treeNode] (document) at (0,0) {};
\node (section1) at (-0.75,-0.5) {\ldots};
\node[treeNode] (section2) at (0,-0.5) {};
\node (section3) at (0.75,-0.5) {\ldots};
\draw[child] (document) to (section1);
\draw[child] (document) to (section2);
\draw[child] (document) to (section3);
\node (subsection1) at (-0.75,-1) {\ldots};
\node[treeNode] (subsection2) at (0,-1) {};
\node (subsection3) at (0.75,-1) {\ldots};
\draw[child] (section2) to (subsection1);
\draw[child] (section2) to (subsection2);
\draw[child] (section2) to (subsection3);
\draw[ultra thick, double, -stealth] (text) to (subsection1);
\node (paragraph1) at (-0.75,-1.5) {\ldots};
\node[treeNode] (paragraph2) at (0,-1.5) {};
\node (paragraph3) at (0.75,-1.5) {\ldots};
\draw[child] (subsection2) to (paragraph1);
\draw[child] (subsection2) to (paragraph2);
\draw[child] (subsection2) to (paragraph3);
\node[treeNode] (sentence1) at (-1.5,-2) {};
\node[treeNode, fill=black] (sentence2) at (-0.75,-2) {};
\node[treeNode] (sentence3) at (0,-2) {};
\node[treeNode] (sentence4) at (0.75,-2) {};
\node[treeNode, fill=black] (sentence5) at (1.5,-2) {};
\draw[child] (paragraph2) to (sentence1);
\draw[child] (paragraph2) to (sentence2);
\draw[child] (paragraph2) to (sentence3);
\draw[child] (paragraph2) to (sentence4);
\draw[child] (paragraph2) to (sentence5);
\node[font=\bfseries, color=blue] (claim) at (0.75,-2.5) {Claim Sentence};
\draw[ultra thick, blue, -stealth] (claim) to (sentence4);
\end{tikzpicture}
\caption{Keyword context of claim sentence: keyword sources in red, the claim sentence in blue.\label{keywordContextFig}}
\end{figure}

Considering keywords in the same sentence is often insufficient. In practice, relevant context is often spread over the entire text. We exploit the structure of the text document in order to collect potentially relevant keywords. Our current implementation uses HTML markup but the document structure could be easily derived from the output format of any word processor. We assume that the document is structured hierarchically into sections, sub-sections etc. For a given claim sentence, we ``walk up'' that hierarchy and add keywords in all headlines we encounter. In addition, we add keywords from the first and preceding sentences in the same paragraph. Figure~\ref{keywordContextFig} illustrates keyword sources for an example claim.

\begin{example}
\label{keywordsExa}
To provide a concrete example, we refer to the paragraph in Figure~\ref{exampleTextFig}. The second sentence contains two claimed results (`three' and `one') that translate into queries of the form: {\scshape SELECT Count(*) FROM T WHERE Games = `indef' AND Category = V}. We illustrate two difficulties associated with these claims.
	
First, there are two claims in one sentence. The system needs to distinguish keywords that are more relevant to each claim. Let's consider the keyword `gambling'. According to the dependency parse tree of the second sentence, the distance from `three' to `gambling' is two while the distance from `one' to `gambling' is one. Then, we assign weights by taking the reciprocal of the distance (see Figure~\ref{exampleKeywordsFig}). This helps the system to understand that `gambling' is more related to `one' than `three'.

Second, no keyword in the second sentence explicitly refers to the restriction {\scshape Games = `indef'}. Rather, it can be implicitly inferred from the context where only the first sentence has the keywords `lifetime bans'. Thereby, considering the keyword context of a claim sentence enables us to identify important and relevant keywords from other parts of the text. In Section~\ref{experimentsSec}, we conduct an experiment to measure the effect of keyword context (see Figure~\ref{keywordsFig}).
\end{example}





\subsection{Constructing Likely Query Candidates}
Having associated both, query fragments and claims, with keywords, we can map claims to likely query candidates. We indexed keyword sets associated with query fragments in an information retrieval engine. For a given claim, we use the associated keyword set to query that information retrieval engine. The returned results correspond to query fragments that are associated with similar keywords as the claim. Furthermore, each returned query fragment is associated with a relevance score, capturing how similar its keywords are to the claim-related keywords. Combining all returned query fragments in all possible ways (within the boundaries of the query model described in Section~\ref{problemSec}) yields the space of claim-specific query candidates. Each candidate is characterized by a single aggregation function fragment, applied to an aggregation column fragment, in the SQL select clause. In addition, each candidate is characterized by a set of unary equality predicates that we connect via a conjunction in the SQL where clause. The SQL from clause can be easily inferred from the other query components: it contains all tables containing any of the columns referred to in aggregates or predicates. We connect those tables via equi-joins along foreign-key-primary-key join paths. 




%% file: sections/model.tex
We map each natural language claim to a probability distribution over matching SQL queries. Based on the most likely query for each claim, we can decide which claims are likely to be wrong and focus the user's attention on those. 


\subsection{Probabilistic Model Overview}
Our probabilistic model is based on a fundamental property of typical text documents (we quantify this effect in Appendix~\ref{testCasesSub}): text summaries tend to have a primary focus. The claims made in a text are not independent from each other but typically connected via a common theme. If we find out the common theme, mapping natural language claims to queries becomes significantly easier. 



We represent the common theme as a document-specific probability distribution over queries. We use that distribution as a prior when inferring likely queries for each claim. Beyond the prior distribution, the likelihood of queries depends on the keyword-based relevance scores that are associated with each claim (we described how those relevance scores can be calculated in the last section). 


\begin{algorithm}[t]
\renewcommand{\algorithmiccomment}[1]{// #1}
\begin{small}
\begin{algorithmic}[1]
\State \Comment{Calculate for each claim in $C$ a distribution over}
\State \Comment{matching queries on database $D$ using relevance}
\State \Comment{scores $S$ via expectation maximization.}
\Function{QueryAndLearn}{$D,C,S$}
\State \Comment{Initialize priors describing text document}
\State $\Theta\gets$\Call{Uniform}{}
\State \Comment{Iterate EM until convergence}
\While{$\Theta$ not converged yet}
\State \Comment{Treat each factual claim}
\For{$c\in C$}
\State \Comment{Calculate keyword-based probability}
\State $Q_c\gets$\Call{TextProbability}{$S,\Theta$}
\EndFor
\State \Comment{Refine probability via query evaluations}
\State $Q\gets$\Call{RefineByEval}{$\{Q_c|c\in C\},C,D$}
\State \Comment{Update document-specific priors}
\State $\Theta\gets$\Call{Maximization}{$\{Q_c|c\in C\}$}
\EndWhile
\State \Return{$\{Q_c|c\in C\}$}
\EndFunction
\end{algorithmic}
\end{small}
\caption{Learn document-specific probability distribution over queries and refine by evaluating query candidates.\label{learningAlg}}
\end{algorithm}

We face a circular dependency: if we had the document theme, we could use it as prior in our search for the most likely query for each claim. On the other side, if we had the most likely query for each claim, we could infer the document theme. This motivates an expectation-maximization approach~\cite{do2008expectation} in which model parameters (describing here the document-specific query distribution) and values of latent variables (describing  claim-specific query distributions) are iteratively refined, using tentative values for one of the two to infer estimates for the other.

Algorithm~\ref{learningAlg} describes how the AggChecker infers probability distributions over query candidates. Starting from a uniform document distribution (captured by parameter $\Theta$ whose precise components are described in the following), the algorithm iterates until convergence. In each iteration, a claim-specific probability distribution over query candidates is calculated for each claim, based on the relevance scores provided as input and the model parameters. 

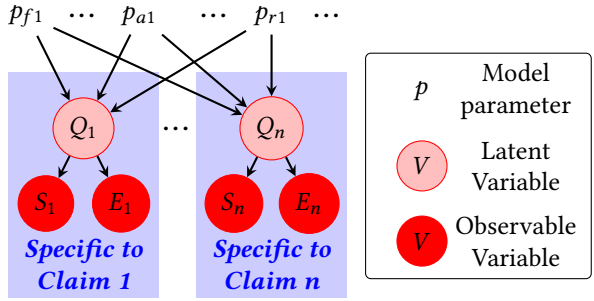
\begin{figure}
\begin{tikzpicture}[modelParameter/.style={font=\bfseries}, RVlatent/.style={fill=red!25, draw=red, circle, align=center}, RV/.style={fill=red, draw=red, circle, align=center}, influence/.style={-stealth,thick}, legendEntry/.style={align=center}]

\node[minimum width=2cm, minimum height=3cm, fill=blue!20] at (-1.25,-2.25) {};
\node[minimum width=2cm, minimum height=3cm, fill=blue!20] at (1.25,-2.25) {};
\node[color=blue, font=\bfseries\itshape, align=center] at (-1.25,-3.3) {Specific to\\Claim 1};
\node[color=blue, font=\bfseries\itshape, align=center] at (1.25,-3.3) {Specific to\\Claim n};

\node[draw=black, minimum width=3cm, minimum height=3cm, rounded corners=2] at (4,-2) {};
\node[modelParameter] at (3.25,-1) {$p$};
\node[legendEntry] at (4.5,-1) {Model\\parameter};
\node[RVlatent] at (3.25,-2) {$V$};
\node[legendEntry] at (4.5,-2) {Latent\\Variable};
\node[RV] at (3.25,-3) {$V$};
\node[legendEntry] at (4.5,-3) {Observable\\Variable};

\node[modelParameter] (pf1) at (-2,0) {$p_{f1}$};
\node[modelParameter] (pf2) at (-1.25,0) {\ldots};
\node[modelParameter] (pa1) at (-0.5,0) {$p_{a1}$};
\node[modelParameter] (pa2) at (0.5,0) {\ldots};
\node[modelParameter] (pr1) at (1.25,0) {$p_{r1}$};
\node[modelParameter] (pr2) at (2,0) {\ldots};

\node[RVlatent] (Q1) at (-1.25,-1.5) {$Q_1$};
\node[RVlatent] (Qn) at (1.25,-1.5) {$Q_n$};
\node at (0,-1.5) {\bfseries\ldots};

\node[RV] (S1) at (-1.75,-2.5) {$S_1$};
\node[RV] (E1) at (-0.75,-2.5) {$E_1$};
\node[RV] (Sn) at (0.75,-2.5) {$S_n$};
\node[RV] (En) at (1.75,-2.5) {$E_n$};

\draw[influence] (pf1) to (Q1);
\draw[influence] (pa1) to (Q1);
\draw[influence] (pr1) to (Q1);

\draw[influence] (pf1) to (Qn);
\draw[influence] (pa1) to (Qn);
\draw[influence] (pr1) to (Qn);

\draw[influence] (Q1) to (S1);
\draw[influence] (Q1) to (E1);
\draw[influence] (Qn) to (Sn);
\draw[influence] (Qn) to (En);
\end{tikzpicture}
\caption{Simplified probabilistic model for query inference: parameters describe prior probabilities of query characteristics, claim queries ($Q_c$) are latent while relevance scores ($S_c$) and evaluation results ($E_c$) are observable.\label{probabilisticModelFig}}
\end{figure}

Strong evidence that a query candidate matches a natural language claim can be obtained by evaluating the query and comparing its result to the claimed one. However, evaluating queries on potentially large data sets may lead to significant processing overheads. Myriads of queries are possible for a given data set and we cannot execute all of them. This is why we use a preliminary, claim-specific query distribution to select promising query candidates for execution. In the next section, we describe efficient processing strategies enabling us to execute hundreds of thousands of query candidates during learning. The evaluation of promising candidates is encapsulated in Function~\textproc{RefineByEval} in Algorithm~\ref{learningAlg}. The result is a refined probability distribution over query candidates for a given claim that takes evaluation results into account. Finally, the model parameters are updated based on the claim-specific distributions. In the following subsections, we provide details on our probabilistic model. A simplified version of the model is illustrated in Figure~\ref{probabilisticModelFig}.

\subsection{Prior Probabilities}
We need to keep that model relatively simple for the following reason: having more parameters to learn typically requires a higher number of iterations until convergence. In our case, each iteration requires expensive data processing and hence we cannot afford an elevated number of steps. We therefore introduce only parameters that describe the probability of certain coarse-grained query features: 
\begin{equation}
\Theta=\langle p_{f1},p_{f2},\ldots,p_{a1},p_{a2},\ldots,p_{r1},p_{r2},\ldots\rangle
\end{equation}

Here, $p_{fi}$ is the prior probability of selecting the $i$-th aggregation function (such as an average or a sum), $p_{ai}$ is the probability of selecting the $i$-th numerical column to aggregate, and $p_{ri}$ is the probability that a restriction (i.e., equality predicate) is placed on the $i$-th column. We can only have a single aggregation function and a single column to aggregate over, the corresponding parameters must therefore sum up to one. This does not apply to the parameters describing the likelihood of a restriction as a query might restrict multiple columns at the same time. During the maximization step of Algorithm~\ref{learningAlg} (line~17), we simply set each component of $\Theta$ to the ratio of maximum likelihood queries with the corresponding property, scaled to the total number of claims in the document. For instance, for updating $r_i$, we divide the number of maximum likelihood queries (summing over all claims) placing a restriction on the $i$-th column by the number of claims. 


\begin{example}
Table~\ref{priorConvergenceTable} shows the convergence of priors $\Theta$ for the example in Figure~\ref{exampleFig}. Note that all three claims in this example have ground truth queries of the form: {\scshape SELECT Count(*) FROM T WHERE Games = `indef' (AND Category = V)}. In Table~\ref{priorConvergenceTable}, priors successfully reflect this pattern after several EM iterations. For instance, the final priors imply the fact that a query with {\scshape Count(*)} is more likely to be the correct one among query candidates. An experimental result in Section~\ref{experimentsSec} demonstrates that the system truly benefits from the prior probabilities (see Table~\ref{probablisticTable}).
\end{example}

\begin{table}
	\centering
	\caption{Changing priors during EM iterations until convergence (for the example in Figure~\ref{exampleFig}).\label{priorConvergenceTable}}
	\begin{small}
		\begin{tabular}{lrrr}
			\toprule[1pt]
			\textbf{Query Fragment} & \textbf{Initial Prior} & \textbf{...} & \textbf{Final Prior}\\
			\midrule[1pt]
			{\scshape Count(*)} & 0.025 & ... & 0.150 \\
			{\scshape Sum(Games)} & 0.025 & ... & 0.012 \\
			... & & & \\
			{\scshape Games =} (any value) & 0.143 & ... & 0.417 \\
			{\scshape Category =} (any value) & 0.143 & ... & 0.297 \\
			... &  &  &  \\
			\bottomrule[1pt]
		\end{tabular}
	\end{small}
\end{table}

\subsection{Claim-specific Probability Distribution}
Next, we show how to calculate claim-specific probability distributions over queries, assuming given values for the parameters above. We introduce random variable $Q_c$ to model the query described in claim $c$. Hence, $\Pr(Q_c=q)$ is the probability that the text for claim $c$ describes a specific query $q$. Variable $Q_c$ depends on another random variable, $S_c$, modeling relevance scores for each query fragment. Those relevance scores are generated by an information retrieval engine, as discussed in the last section, based on keywords in claim text. If a query fragment has high relevance scores (i.e., many related keywords appear in the claim), the probability of $Q_c$ for queries containing that fragment increases. Also, if a query evaluates to the claimed result, the probability that the claim describes this query should intuitively increase. We model by $E_c$ evaluation results for a set of promising query candidates and $Q_c$ depends on $E_c$. According to the Bayes rule, we obtain:
\begin{equation}
\Pr(Q_c|S_c,E_c)\propto \Pr(S_c\wedge E_c|Q_c)\cdot\Pr(Q_c)
\end{equation}



$\Pr(Q_c)$ denotes prior query probabilities, derived from a document-specific theme. Assuming independence between relevance scores and evaluation results, we obtain:
\begin{equation}
\Pr(S_c\wedge E_c|Q_c)=\Pr(S_c|Q_c)\cdot \Pr(E_c|Q_c)
\end{equation}

We assume independence between different query characteristics. This is a simplification (as certain aggregation functions might often be used with certain aggregation columns for instance), but modeling dependencies would require additional parameters and our prior remark about model complexity applies. Via independence assumptions, we obtain:
\begin{equation}
\Pr(S_c|Q_c)=\Pr(S_c^F|Q_c)\cdot\Pr(S_c^A|Q_c)\cdot\Pr(S_c^R|Q_c)
\end{equation}

Variable $S_c^F$ represents the relevance scores assigned to each aggregation function by the information retrieval engine, based on keywords surrounding claim $c$. By $S_c^A$, we denote relevance scores for aggregation columns, and by $S_c^R$ scores for query fragments representing restrictions (i.e., equality predicates). The probability $\Pr(S_c^A|Q_c)$ is for instance the probability that we obtain relevance scores $S_c^A$, assuming that the text author describes query $Q_c$ in claim $c$. Obtaining a high relevance score for a query fragment (be it aggregation function, column, or predicate) means that related keywords appear prominently in the claim text. Hence, query fragments that are part of the claim query should tend to receive higher relevance scores than the others.

We use a simple model that complies with the latter intuition: the probability to receive certain relevance scores is proportional to the relevance scores of the fragments appearing in the claim query. E.g., assume that claim query $q$ aggregates over column $a\in A$ (where $A$ designates the set of all candidate columns for aggregates in the database). We denote by $S_c^A(a)$ the relevance score for the query fragment representing that specific column $a$. We set $\Pr(S_c^A|Q_c=q)=S_c^A(a)/\sum_{a\in A}S_c^A(a)$, scaling relevance scores to the sum of relevance scores over all query fragments in the same category (i.e., aggregation columns in this case).


Correct claims are more likely than incorrect claims in typical text documents. Hence, if a query candidate evaluates to the claimed value, it is more likely to be the query matching the surrounding text. It is typically not feasible to evaluate all candidate queries on a database. Hence, we restrict ourselves to evaluating queries that have high probabilities based on relevance scores alone (line~14 in Algorithm~\ref{learningAlg}). Let $E_c$ be the evaluation results of promising queries for claim $c$. We set $\Pr(E_c|Q_c=q)=p_T$ if the evaluations $E_c$ map query $q$ to a result that rounds to the claimed value. We set $\Pr(E_c|Q_c=q)=1-p_T$ otherwise. Parameter $p_T$ is hence the assumed probability of encountering true claims in a document. Different settings realize different tradeoffs between precision and recall (see Section~\ref{experimentsSec}). 

We finally calculate prior query probabilities, assuming again independence between different query characteristics:
\begin{equation}
\Pr(Q_c)=\Pr(Q_c^F)\cdot\Pr(Q_c^A)\cdot\prod_{r\in R}\Pr(Q_c^r)
\end{equation}

The prior probabilities of specific aggregation functions ($\Pr(Q_c^F)$), aggregation columns ($\Pr(Q_c^A)$), and restrictions ($\Pr(Q_c^r)$) follow immediately from the model parameters $\Theta$. In summary, let $q$ be an SQL query with aggregation function $f_q$ and aggregation column $a_q$, containing equality predicates restricting the $i$-th column to value $V_q(i)$ (with $V_q(i)=*$ if no restriction is placed on the $i$-th column). The probability $\Pr(Q_c=q)$ that query $q$ is described in claim $c$ is \textit{proportional} to the product of the following factors: the prior probability of $q$ appearing in the current document (i.e., $p_{f_q}\cdot p_{f_a}\cdot \prod_{i:V_q(i)\neq *}p_{r_i}$), the likelihood to receive the observed keyword-based relevance scores for $q$'s query fragments (i.e., $S_c(f_q)\cdot S_c(a_q)\cdot\prod_{i}S_c(r_i=V_q(i))$ where $S_c()$ generally maps fragments to their scores for claim $c$), and $p_T$ if the rounded query result matches the claim value (or $1-p_T$ otherwise). We can use this formula to map each claim to a maximum likelihood query (line~12 in Algorithm~\ref{learningAlg}). Note that it is not necessary to scale relevance scores since the scaling factor is constant for each specific claim. We only compare query candidates that have been selected for evaluation based on their keyword and prior-based probabilities alone (line~15 in Algorithm~\ref{learningAlg}). Before providing further details on query evaluation, we present the following example to elaborate on the benefit of \textproc{RefineByEval} at line~15 in Algorithm~\ref{learningAlg}.

\begin{example}
Let's again use the example with claimed result `one' introduced in Figure~\ref{exampleFig}. Without near-perfect natural language understanding, it is almost impossible to map the phrase \textit{``lifetime bans''} to query fragment {\scshape Games = `indef'}. Nevertheless, the learned priors during EM iterations will at least tell us that a restriction is usually placed on column {\scshape Games} (11 out of 13 claims in this article~\cite{53814-2} have ground truth queries with restriction on {\scshape Games}). By evaluating many related query candidates, the system can find out that {\scshape SELECT Count(*) FROM nflsuspensions WHERE Games = `indef' AND Category = `gambling'} yields the same result as claimed in text (and largely no other queries do). Since we boost the probability of queries that evaluate to the claimed value, this query gets a higher refined probability (see Figure~\ref{exampleProbabilityFig}). It is noteworthy to mention that this is possible only when the system has learned the correct priors reflecting the common theme of ground truth queries. Nevertheless, articles typically have some easy cases (i.e., claims with distinctive keywords nearby) where the system can correctly translate into queries. Then, the system can also cope with other more difficult cases as the information gained from easy cases spreads across claims through EM iterations. Thus, the system benefits from easy cases and successfully learns the correct priors. In summary, the overall effect of \textproc{RefineByEval} on our test cases is presented in Section~\ref{experimentsSec} (see Table~\ref{probablisticTable}).
\end{example}





%% file: sections/query.tex
\begin{algorithm}[t]
\renewcommand{\algorithmiccomment}[1]{// #1}
\begin{small}
\begin{algorithmic}[1]
\State \Comment{Calculate aggregates $S_{FA}$ on data $D$ for row sets}
\State \Comment{defined by predicates on columns $G$ with non-zero}
\State \Comment{probability according to query distribution $Q$.}
\Function{Cube}{$Q,D,G,S_{FA}$}
\State \Comment{Collect directly referenced tables}
\State $T\gets \{$\Call{Table}{$x$}$|\langle f,x\rangle\in S_{FA}\vee x\in D\}$
\State \Comment{Add tables and predicates on join paths}
\State $T\gets T\cup\{$\Call{JoinPathTables}{$t_1,t_2$}$|t_1,t_2\in T\}$
\State $J\gets\{$\Call{JoinPathPreds}{$t_1,t_2$}$|t_1,t_2\in T\}$
\State \Comment{Collect relevant literals for \textit{any} claim}
\State $L\gets\{$\Call{Literals}{$r$}$|r\in D\wedge \exists c\in C:\Pr(l|Q_c)>0\}$
\State \textbf{return} \textproc{ExecuteQuery}$(D,$ ``
\Statex \hspace{1cm} {\scshape SELECT} $S_{FA}, D$ {\scshape FROM} 
\Statex \hspace{1.5cm} {\scshape ( SELECT} $S_{FA}, $\Call{InOrDefault}{$P,L$} 
\Statex \hspace{1.5cm} {\scshape FROM} $T$ {\scshape WHERE} $J$ {\scshape) CUBE BY} $P$ ''$)$
\EndFunction
\vspace{0.15cm}
\State \Comment{Refine probabilities $Q$ for claims $C$ on data $D$.}
\Procedure{RefineByEval}{$Q,C,D$}
\State \Comment{Evaluate likely queries for each claim}
\For{$c\in C$}
\State \Comment{Pick scope for query evaluations}
\State $\langle S_F,S_A,S_R\rangle\gets$\Call{PickScope}{$Q,C,D$}
\State \Comment{Initialize query evaluation results}
\State $E\gets\emptyset$
\State \Comment{Iterate over predicate column group}
\For{$G\subseteq S_R:|G|=n_G(|S_R|)$}
\State \Comment{Form likely aggregates}
\State $S_{FA}\gets F\times A$
\State \Comment{Exploit cache content}
\For{$fa\in S_{FA}|$\Call{IsCached}{$fa,G$}}
\State $E\gets E\cup$\Call{CacheGet}{$fa,G$}
\State $S_{FA}\gets S_{FA}\setminus \{fa\}$
\EndFor
\State \Comment{Generate missing results}
\State $E\gets E\cup$\Call{Cube}{$Q,D,G,S_{FA}$}
\State \Comment{Store in cache for reuse}
\State \Call{CachePut}{$S_{FA},G,E$}
\EndFor
\State \Comment{Refine query probabilities}
\State $Q_c\gets$\Call{Refine}{$Q_c,E$}
\EndFor
\State \Return{$Q$}
\EndProcedure
\end{algorithmic}
\end{small}
\caption{Refine query probabilities by evaluations. \label{executionAlg}} 
\end{algorithm}

In fact-checking, we can partially resolve ambiguities in natural language understanding by evaluating large numbers of query candidates. As we show in Section~\ref{experimentsSec}, the ability to process large numbers of query candidates efficiently turns out to be crucial to make fact-checking practical. 


The expectation maximization approach presented in the previous section relies on a sub-function, \textproc{RefineByEval}, to refine query probabilities by evaluating candidates. Algorithm~\ref{executionAlg} implements that function. The input is an unrefined probability distribution over query candidates per natural language claim, as well as the claim set and the database they refer to. The output is a refined probability distribution, taking into account query evaluation results.

\subsection{Prioritizing Query Evaluations}
Ideally, we would be able to evaluate all query candidates with non-zero probability based on the initial estimates. By matching the results of those queries to results claimed in text, we would be able to gain the maximal amount of information. In practice, we must select a subset of query candidates to evaluate. To make the choice, we consider two factors. First, we consider the a-priori likelihood of the query to match the claim. Second, we consider the processing efficiency of evaluating many queries together.

Queries are characterized by three primary components in our model: an aggregation function, an aggregation column, and a set of predicates. In a first step (line~19 in Algorithm~\ref{executionAlg}), we determine the evaluation scope as the set of alternatives that we consider for each query characteristic. To determine the scope, we use a cost model that takes into account the size of the database as well as the number of claims to verify. Function~\textproc{PickScope} exploits the marginal probabilities of query characteristics. It expands the scope, prioritizing more likely alternatives, until estimated evaluation cost according to our cost model reaches a threshold.



\subsection{Merging Query Candidates}
As a naive approach, we could form all query candidates within the scope and evaluate them separately. This would however neglect several opportunities to save computation time by avoiding redundant work. First, alternative query candidates for the same claim tend to be quite similar. This means that the associated query plans can often share intermediate results, thereby amortizing computational overheads. Second, even the query candidates for different claims in the same document tend to be similar, thereby enabling us to amortize cost again. This relates to our observation (quantified in Section~\ref{experimentsSec}) that different claims in the same document are often semantically related. Finally, Algorithm~\ref{executionAlg} will be called repeatedly for the same document (over different iterations of the expectation maximization approach). In particular in later iterations, topic priors change slowly and likely query candidates for the same claim change only occasionally between iterations. This enables us to reuse results from previous iterations. Algorithm~\ref{executionAlg} exploits all three opportunities to avoid redundant work.

At the lowest layer, we use cube queries to efficiently calculate aggregates for different data subsets. Each data subset is associated with one specific combination of query predicates. One cube query can therefore cover many alternative query candidates as long as their equality predicates refer to a small set of columns (the cube dimensions). Executing a cube query on a base table yields aggregates for each possible value combination in the cube dimension. The result set can be large if cube dimensions contain many distinct values. In the context of fact-checking, we are typically only interested in a small subset of values (the ones with non-zero marginal probabilities, meaning that corresponding matches are returned after keyword matching). We can however not filter the data to rows containing those values before applying the cube operator: this would prevent us from obtaining results for query candidates that do not place any predicate on at least one of the cube dimensions. Instead, we apply the cube operator to the result of a sub-query that replaces all literals with zero marginal probability by a default value (function~\textproc{InOrDefault}). This reduces the result set size while still allowing us to evaluate all related query candidates. Note that evaluating multiple aggregates for the same cube dimensions in the same query is typically more efficient than evaluating one cube query for each aggregate. Hence, we merge as many aggregates as possible for the same dimensions into the same query.

\begin{example}
All four query candidates that are explicitly shown in Figure~\ref{exampleProbabilityFig} have the form: {\scshape SELECT Fct(Agg) FROM T WHERE Category = V (AND Games = `indef')}. Thus, it is sufficient to evaluate just one cube query to get their results: {\scshape SELECT Count(*), Percentage(Category), Sum(Games) FROM T GROUP BY CUBE(Category, Games)}. Nevertheless, we can notice that its result set would also contain many unnecessary aggregates for every distinct value combination of columns {\scshape Category} and {\scshape Games}. We instead restrict the cube query to only a small subset of values of interest:

{\scshape SELECT Count(*), Percentage(Category), Sum(Games), CASE WHEN Category = `gambling' THEN 1 WHEN Category = `substance abuse' THEN 2 ELSE -1 END AS InOrDefaultCategory, CASE WHEN Games = `indef' THEN 1 ELSE -1 END AS InOrDefaultGames FROM T GROUP BY CUBE(InOrDefaultCategory, InOrDefaultGames)}.

\end{example}

\subsection{Caching across Claims and Iterations}
Furthermore, we avoid redundant computation by the use of a cache. The cache is accessed via functions \textproc{IsCached}, \textproc{CacheGet}, and \textproc{CachePut} with the obvious semantics. The cache persists across multiple iterations of the main loop in Algorithm~\ref{executionAlg} and across multiple invocations of Algorithm~\ref{executionAlg} for the same document (during expectation maximization). We avoid query evaluations if results are available in the cache and cache each generated result. We can choose the granularity at which results are indexed by the cache. Indexing results at a coarse granularity might force us to retrieve large amount of irrelevant data. Indexing results at a very fine granularity might create overheads when querying the cache and merging result parts. We found the following granularity to yield a good performance tradeoff: we index (partial) cube query results by a combination of one aggregation column, one aggregation function, and a set of cube dimensions. The index key does not integrate the set of relevant literals in the cube columns, although the set of literals with non-zero marginal probability may vary across different claims or iterations. We generate results for all literals that are assigned to a non-zero probability for any claim in the document (this set is equivalent to the set of literals in predicates returned during keyword matching). This simplifies indexing and is motivated by the observation that different claims tend to have high overlap in the sets of relevant literals for a given column. 

To create more opportunities to share partial results, we cover the query scope via multiple cube queries, iterating over subsets of cube dimensions. Additionally, this prevents us from generating results for cube queries with an unrealistically high number of cube dimensions (e.g., we expect at most three predicates per claim in typical newspaper articles while the query scope may include several tens of predicate columns). On the other hand, we increase the total number of queries and generate redundant results. We use function $n_G(x)$ to pick the number of dimensions for each cube query. We chose $n_G(x)=\max(m,x-1)$ for our Postgres-based cube operator implementation where $m$ is the maximal number of predicates per claim (we use $m=3$). 
Function~\textproc{Cube} in Algorithm~\ref{executionAlg} constructs the cube query (we use simplified SQL in the pseudo-code), executes it, and returns the result. It uses function~\textproc{Table} to retrieve associated database tables for aggregates and predicates. It exploits join paths to identify connecting tables and join predicates. Our approach assumes that the database schema is acyclic.

%% file: sections/experiments2.tex
\pgfplotscreateplotcyclelist{patternList}{%
	{fill=blue!25, postaction={pattern=horizontal lines}},
	{fill=red!50, postaction={pattern=north east lines}},
	{fill=green, postaction={pattern=vertical lines}},
	{fill=yellow, postaction={pattern=north west lines}},
	{fill=cyan, postaction={pattern=dots}}
}

\pgfplotscreateplotcyclelist{markerList}{%
	{blue, mark=*},
	{red, mark=x},
	{green, mark=diamond}
}

We evaluated the AggChecker on 53 real articles, summarizing data sets and featuring 392 claims. Those test cases range from New York Times and 538 newspaper articles to summaries of Internet surveys. Test case statistics and details on the test case collection process can be found in Appendix~\ref{testCasesSub}. Using our tool, we were able to identify multiple erroneous claims in those articles, as confirmed by the article authors.

\subsection{Evaluation Metrics}
\label{metricsSub}

\textit{F1 score} measures a system's performance on identifying erroneous claims. We calculate the F1 score based on these definitions of \textit{precision} and \textit{recall}:
\begin{definition}\textbf{Precision} is the fraction of truly erroneous (according to ground truth) claims that the system has tentatively marked as erroneous.
\end{definition}
\begin{definition}\textbf{Recall} is the fraction of claims identified by the system as erroneous among the total set of truly erroneous claims.
\end{definition}

\subsection{User Study}
\label{userStudySub}


We performed an anonymized user study to see whether the AggChecker enables users to verify claims more efficiently than generic interfaces. We selected six articles from diverse sources (538, the New York Times, and Stack Overflow). We selected two long articles featuring more than 15 claims~\cite{Stackoverflow16, 53814} and four shorter articles with five to ten claims each~\cite{Stackoverflow15, NYT14, 53815, 53816}. Users had to verify claims in those documents, either by executing SQL queries on the associated database or by using the AggChecker. We gave a time limit of 20 minutes for each of the two longer articles and five minutes for each of the shorter ones. Users were alternating between tools and never verified the same document twice. We had eight participants, seven of which were computer science majors. We gave users a six minute tutorial for AggChecker. 


\begin{figure}
\center
\begin{tikzpicture}[mark size=1pt]
\begin{groupplot}[width=3.75cm, height=3cm, %
group style={group size = 3 by 2, x descriptions at=edge bottom, ylabels at=edge left, horizontal sep=0.3cm, vertical sep=0.2cm}, %
ymin=0, xlabel={Time (s)}, ylabel={\#Verified}, ylabel near ticks, ymajorgrids, xmajorgrids, legend to name=userMainLegend, cycle list name=markerList]
\nextgroupplot[title={}, mark repeat=50]
\addplot table[x index=0, y index=7, col sep=comma] {plots_used/userStudy_StackOverflow2016.txt};
\addplot table[x index=0, y index=8, col sep=comma] {plots_used/userStudy_StackOverflow2016.txt};

\nextgroupplot[title={}, mark repeat=10]
\addplot table[x index=0, y index=9, col sep=comma] {plots_used/userStudy_StackOverflow2015.txt};
\addplot table[x index=0, y index=10, col sep=comma] {plots_used/userStudy_StackOverflow2015.txt};

\nextgroupplot[title={}, mark repeat=10]
\addplot table[x index=0, y index=9, col sep=comma] {plots_used/userStudy_eloblatter.txt};
\addplot table[x index=0, y index=10, col sep=comma] {plots_used/userStudy_eloblatter.txt};

\nextgroupplot[title={}, mark repeat=50]
\addplot table[x index=0, y index=7, col sep=comma] {plots_used/userStudy_flyingetiquettesurvey.txt};
\addplot table[x index=0, y index=8, col sep=comma] {plots_used/userStudy_flyingetiquettesurvey.txt};

\nextgroupplot[title={}, mark repeat=10]
\addplot table[x index=0, y index=8, col sep=comma] {plots_used/userStudy_SundayShows.txt};
\addplot table[x index=0, y index=9, col sep=comma] {plots_used/userStudy_SundayShows.txt};

\nextgroupplot[title={}, mark repeat=10, legend entries={AggChecker (Avg), SQL (Avg)}, legend columns=2]
\addplot table[x index=0, y index=9, col sep=comma] {plots_used/userStudy_hiphopcandidatelyrics.txt};
\addplot table[x index=0, y index=10, col sep=comma] {plots_used/userStudy_hiphopcandidatelyrics.txt};
\end{groupplot}
\end{tikzpicture}

\ref{userMainLegend}
\caption{Number of correctly verified claims as a function of time, using different tools on different articles. Articles left-to-right then top-to-bottom: \protect\cite{Stackoverflow16}, \protect\cite{Stackoverflow15}, \protect\cite{53815}, \protect\cite{53814}, \protect\cite{NYT14}, \protect\cite{53816}.\label{userMainFig}
}
\end{figure}
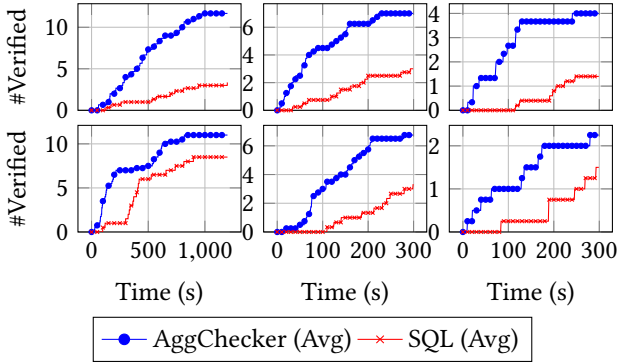

\begin{figure}[t]
\center
\begin{tikzpicture}[mark size=1pt]
\begin{groupplot}[width=4.5cm, height=3cm, %
group style={group size = 2 by 1, x descriptions at=edge bottom, ylabels at=edge left, horizontal sep=0.4cm}, %
ymin=0, ylabel={\#Verified/Min}, ylabel near ticks, ymajorgrids, legend to name=userThroughputLegend, ybar=0pt, xtick=\empty, x label style={at={(0.5,0.3)}}, legend columns=2, cycle list name=patternList]
\nextgroupplot[bar width=3pt, xlabel={User}, xticklabels={}, xmin=0.5, xmax=8.5]
\addplot table[x expr=1+\coordindex, y index=1, col sep=comma] {plots_used/userStudyPerMinuteByUser.txt};
\addplot table[x expr=1+\coordindex, y index=2, col sep=comma] {plots_used/userStudyPerMinuteByUser.txt};
\nextgroupplot[bar width=3pt, xlabel={Article}, xticklabels={}, xticklabel style={rotate=90}, legend entries={AggChecker, SQL}]
\addplot table[x expr=1+\coordindex, y index=1, col sep=comma] {plots_used/userStudyPerMinuteByArticle.txt};
\addplot table[x expr=1+\coordindex, y index=2, col sep=comma] {plots_used/userStudyPerMinuteByArticle.txt};
\end{groupplot}
\end{tikzpicture}

\ref{userThroughputLegend}
\caption{Number of claims verified per minute, grouped by user and by article. Articles left to right: \protect\cite{Stackoverflow16}, \protect\cite{53814}, \protect\cite{Stackoverflow15}, \protect\cite{53815}, \protect\cite{NYT14}, \protect\cite{53816}.
\label{userThroughputFig}}
\end{figure}
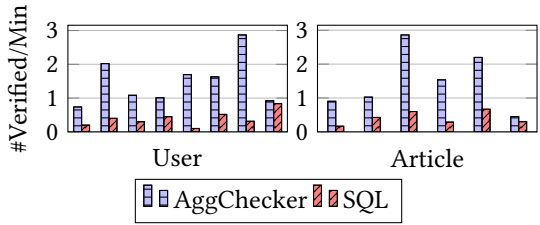

\begin{table}[t]
	\centering
	\caption{Verification by used AggChecker features.\label{featuresTable}}
	\begin{small}
		\begin{tabular}{rrrr}
			\toprule[1pt]
			\textbf{Top-1} & \textbf{Top-5} & \textbf{Top-10} & \textbf{Custom} \\
			(1 click) & (2 clicks) & (3 clicks) & \\
			\midrule[1pt]
			44.5\% & 38.1\% & 4.6\% & 12.8\% \\
			\bottomrule[1pt]
		\end{tabular}
	\end{small}
\end{table}

\begin{table}[t]
	\centering
	\caption{Results of on-site user study.\label{methodsTable}}
	\begin{small}
		\begin{tabular}{lrrr}
			\toprule[1pt]
			\textbf{Tool} & \textbf{Recall} & \textbf{Precision} & \textbf{F1 Score} \\
			\midrule[1pt]
			AggChecker + User & 100.0\% & 91.4\% & 95.5\% \\
			\midrule
			SQL + User & 30.0\% & 56.7\% & 39.2\% \\
			\bottomrule[1pt]
		\end{tabular}
	\end{small}
\end{table}

Figure~\ref{userMainFig} reports the number of correctly verified claims for different articles and tools as a function of time. We collected timestamps from the AggChecker interface and analyzed SQL logs to calculate times for SQL queries. We count a claim as verified if the user selected the right query in the AggChecker interface or if the right query was issued via SQL. Clearly, users tend to be more efficient when using the AggChecker in each single case. Figure~\ref{userThroughputFig} aggregates those results into the ``fact-checking throughput'', meaning the number of correctly verified claims per minute, grouped by user (left) and by article (right). It turns out that users are in average six times faster at verifying claims when using the AggChecker interface. Table~\ref{featuresTable} shows that this speedup is mainly due to our ability to automatically map claims to the right query in most cases (in 82.6\% of cases, users selected one of the top-5 proposed queries). 

Table~\ref{methodsTable} finally adopts a complementary perspective and measures the number of erroneous claims that was found (recall) and the percentage of incorrect claims among the ones marked up as incorrect (precision). We identified in total three erroneous claims in our six test case articles. As expected, users achieve highest scores when interacting with the AggChecker. In a final survey, five out of eight users had a ``much better'' overall experience with the AggChecker while the remaining three had a ``better'' experience. 






\subsection{Automated Checking Performance}
\label{baselinesSub}

Table~\ref{baselineComparisonTable} compares all baselines in terms of precision, recall, F1 score, and run time. We also compare different variants of the AggChecker. We simplify, in multiple steps, different parts of the system and evaluate the performance. The first section of Table 5 analyzes the impact of the keyword context. In our main AggChecker version, we consider keywords from multiple sources beyond the claim sentence: keywords from the previous sentence, the first sentence in a paragraph, we consider synonyms of other keywords, and keywords from preceding headlines. In Table 5, we add those keyword sources successively, resulting in significant improvements in precision and F1 score. Next, we consider simplifications to our probabilistic model. Our main version considers random variables associated with keyword-based relevance scores ($S_c$), query evaluation results ($E_c$), and priors ($\Theta$). We add those variables step by step, demonstrating significant improvements in F1 scores. Any simplification of our probabilistic model therefore worsens performance. Next, we analyze tradeoffs between processing time and result quality. We vary the number of query fragments, retrieved via Lucene, that are considered during processing. Considering higher number of fragments increases the F1 score but leads to higher processing overheads. We use 20 query fragments per claim in our main AggChecker version, realizing what we believe is the most desirable tradeoff between result quality and processing overheads. Altogether, our results demonstrate that each component of our system is important in order to achieve best performance. We executed a second series of experiments, comparing different variants according to different metrics, which are shown in Appendix~\ref{automatedSub}.

Next, we compare against another system that focuses on automated fact checking. ClaimBuster~\cite{HassanA17, HassanZ17} is a recently proposed system for automated fact-checking of text documents. ClaimBuster supports a broader class of claims while we focus on numerical aggregates. We demonstrate in the following that this specialization is necessary to achieve good performance for the claim types we consider. ClaimBuster comes in multiple versions. ClaimBuster-FM matches input text against a database containing manually verified statements with truth values. ClaimBuster-FM returns the most similar statements from the database, together with similarity scores. We tried aggregating the truth values of the returned matches in two ways: ClaimBuster-FM (Max) uses the truth value of the statement with maximal similarity as final result. ClaimBuster-FM (MV) uses the weighted majority vote, weighting the truth value of each returned statement by its similarity score. 

\begin{table}
	\centering
	\caption{Comparison of AggChecker with baselines.\label{baselineComparisonTable}}
	\begin{small}
		\begin{tabular}{p{2.9cm}rrrr}
			\toprule[1pt]
			\textbf{Tool} & \textbf{Recall} & \textbf{Precision} & \textbf{F1} & \textbf{Time} \\
			\midrule[1pt]
			\multicolumn{5}{c}{\textbf{AggChecker - Keyword Context} (Figure~\ref{keywordsFig})} \\
			Claim sentence & 70.8\% & 29.3\% & 41.7\% & - \\
			+ Previous sentence & 68.8\% & 31.1\% & 42.9\% & - \\
			+ Paragraph Start & 70.8\% & 31.8\% & 43.9\% & - \\
			+ Synonyms & 70.8\% & 34.3\% & 46.3\% & - \\
			+ Headlines (current version) & 70.8\% & 36.2\% & \textbf{47.9\%} & - \\
			\midrule
			\multicolumn{5}{c}{\textbf{AggChecker - Probabilistic Model} (Table~\ref{probablisticTable})} \\
			Relevance scores $S_c$ & 93.8\% & 13.3\% & 23.3\% & - \\
			+ Evaluation results $E_c$ & 70.8\% & 32.7\% & 44.7\% & - \\
			+ Learning priors $\Theta$ (current version) & 70.8\% & 36.2\% & \textbf{47.9\%} & - \\
			\midrule
			\multicolumn{5}{c}{\textbf{AggChecker - Time Budget by Lucene Hits} (Figure~\ref{tradeoffFig})} \\
			\# Hits = 1 & 79.2\% & 20.1\% & 32.1\% & 108s \\
			\# Hits = 10 & 70.8\% & 33.7\% & 45.6\% & 121s \\
			\# Hits = 20 (current version) & 70.8\% & 36.2\% & \textbf{47.9\%} & 128s \\
			\# Hits = 30 & 68.8\% & 36.3\% & 47.5\% & 133s \\
			\midrule
			\multicolumn{5}{c}{\textbf{Baselines}} \\
			ClaimBuster-FM (Max) & 34.1\% & 12.3\% & 18.1\% & 142s \\
			ClaimBuster-FM (MV) & 31.7\% & 15.9\% & 21.1\% & 142s \\
			ClaimBuster-KB + NaLIR & 2.4\% & 10.0\% & 3.9\% & 18733s \\
			AggChecker Automatic & 70.8\% & 36.2\% & \textbf{47.9}\% & \textbf{128s} \\
			\bottomrule[1pt]
		\end{tabular}
	\end{small}
\end{table}

Another version of ClaimBuster (ClaimBuster-KB) transforms an input statement into a series of questions, generated by a question generation tool~\cite{HeilmanS09, HeilmanS10}. Those questions are sent as queries to knowledge bases with natural language interfaces (e.g, Wolfram Alpha and Google Answers). The bottleneck however is that the required data for our test cases is not available in these generic knowledge bases. Nevertheless, a natural language query interface running on a database with all our data sets can be used as an alternative knowledge base for ClaimBuster-KB instead. To do so, we use NaLIR~\cite{LiJ14, Li2014}, a recently proposed natural language database query interface. We cannot directly use NaLIR for fact-checking as its input format only allows natural language queries (not claims). Thus, we use the same question generation tool as ClaimBuster-KB to transform claims into question sequences and send them (including the original sentence) as queries to NaLIR. Then, we compare the results from NaLIR with the claimed value in text to see if there is a match on at least one of the queries. If so, we verify the claim as correct and if not, as wrong. Note that, using the original code of NaLIR, less than 5\% of sentences are translated into SQL queries while others throw exceptions during the translation process. We extended the code to support a broader range of natural language queries (e.g., by implementing a more flexible method for identifying command tokens in parse trees) which increased the translation ratio to 42.1\%. Still, only 13.6\% of the translated queries return a single numerical value which can be compared with the claimed value in text.

\begin{table}
	\centering
	\caption{Run time for all test cases.\label{totalTimeTable}}
	\begin{small}
		\begin{tabular}{lrrl}
			\toprule[1pt]
			\textbf{Version} & \textbf{Total (s)} & \textbf{Query (s)} & \textbf{Speedup}\\
			\midrule[1pt]
			Naive & 2587 & 2415 & -\\
			\midrule
			+ Query Merging & 151 & 39 & $\times61.9$ \\
			\midrule
			+ Caching & 128 & 18 & $\times2.1$ \\
			\bottomrule[1pt]
		\end{tabular}
	\end{small}
\end{table}

In Table~\ref{baselineComparisonTable}, the AggChecker outperforms the other baselines by a significant margin. The reasons vary across baselines. ClaimBuster-FM relies on manually verified facts in a repository. This covers popular claims (e.g., made by politicians) but does not cover ``long tail'' claims. We verified that the relatively high recall rate of ClaimBuster-FM is in fact due to spurious matches, the necessary data to verify those claims is not available.

Prior work on natural language query interfaces has rather focused on translating relatively concise questions that a user may ask. The claim sentences in our test data tend to be rather complex (e.g., multi-line sentences with multiple sentence parts). This makes it already hard to derive relevant questions for verification. Also, sentence parse tree and query tree tend to have a high edit distance (which hurts approaches such as NaLIR that assume high similarity). Further, many claims (30\%) do not explicitly state the aggregation function (in particular for counts and sums) or there are multiple claims within the same sentence (29\%). All of those challenges motivate specialized approaches for fact-checking from raw relational data.

\textbf{Effect of massive processing.}
We evaluate a large number of query candidates to refine verification accuracy. An efficient processing strategy is required to avoid excessive computational overheads. Table~\ref{totalTimeTable} demonstrates the impact of the optimizations discussed in Section~\ref{querySec} (we used a laptop with 16 GB RAM and a 2.5 GHZ Intel i5-7200U CPU running Windows 10). We report total execution times and also only the query processing times for fact-checking our entire set of test cases. Processing candidates naively yields query processing times of more than 40 minutes. Merging query candidates and caching query results yield accumulated processing time speedups of more than factor 129.9.

%% file: sections/related.tex
\begin{table}
	\centering
	\caption{Properties of AggChecker and ClaimBuster.\label{claimbusterComparisonTable}}
	\begin{small}
		\begin{tabular}{lp{3cm}p{3cm}}
			\toprule[1pt]
			& \textbf{AggChecker} & \textbf{ClaimBuster} \\
			\midrule[1pt]
			\textbf{Task} & Fact-checking & Claim identification, \\
			& & Fact-checking \\
			\midrule
			\textbf{Claims} & Numerical & Generic \\
			\midrule
			\textbf{Data} & Structured & Unstructured, Structured \\
			\midrule[1pt]
			\textbf{Summary} & Specialized to claims on numerical aggregates & Broader task and claim scope \\
			\bottomrule[1pt]
		\end{tabular}
	\end{small}
\end{table}

ClaimBuster~\cite{HassanA17, HassanZ17} supports users in fact-checking natural language texts. ClaimBuster verifies facts by exploiting natural language fact checks prepared by human fact checkers, natural language query interfaces to existing knowledge bases, or textual Google Web query results. In short, Table~\ref{claimbusterComparisonTable} gives a comparison of the two systems.

We focus on facts (i.e., numerical aggregates) that are not explicitly given but can be derived from the input data set. Most prior work on fact-checking~\cite{NakasholeM14, VlachosR14, VlachosR15, CiampagliaS15, ShiW15, ShiW16, HassanZ17, HassanA17, Wang17, ThorneV17, KaradzhovN17} assumes that entities a claim refers to are readily available in a knowledge base~\cite{BabakarM16}. Prior work on argument mining~\cite{PeldszusS13, LippiT16-1, LippiT16-2} identifies complex claims and supporting arguments in text. We link text claims to structured data instead. Prior work tests the robustness of claim queries against small perturbations~\cite{WuA14, WuW14, WalenzY16, WuA17}. Techniques such as query merging and caching can be used in this context as well. However, as the claim SQL query is given as input, this line of work avoids the primary challenge that we address: the translation of natural language claims into SQL queries. The problem of translating natural language queries or keyword sets to SQL queries has received significant attention~\cite{Agrawal2002, LiJ14, Li2014, Saha2016}. Fact-checking differs as users specify queries together with claimed results, allowing new approaches. Also, we are operating on entire documents instead of single queries.

%% file: sections/conclusion.tex
We introduced the problem of fact-checking natural language summaries of relational databases. We have presented a first corresponding approach, encapsulated into a novel tool called the AggChecker. We successfully used it to identify erroneous claims in articles from major newspapers. 

%% file: sections/appendix3.tex
\section{User Preferences}




We report on the results of a survey that we conducted among the participants of the user study described in Section~\ref{userStudySub}. Table~\ref{detailedSurveyTable} summarizes the survey results. We asked users for preferences when comparing SQL to the AggChecker interface. Our scale ranges from a strong preference for SQL (represented a \textbf{SQL++} in Table~\ref{detailedSurveyTable}), over equal preferences ($\mathbf{SQL\approx AC}$), up to a strong preference for the AggChecker (\textbf{AC++}). We also asked users to compare tools according to specific categories. Namely, we asked about learning overheads, preference for verifying correct claims, and for verifying incorrect claims. 




We also collected free form comments from our users which were generally very favorable. A few examples follow. ``\textit{Shortened time considerably. Very Very friendly to non-cs majors. Interface is very simple and easy to follow. I was able to finish all the ones that I used Face Checker before the time ran out so it's awesome. If I had to choose between SQL and Face Checker, I'd choose Fact Checker any day.}`` ``\textit{The suggestions made it very easy to create custom queries.}'' 

\begin{table}
	\center
	\caption{Results of user survey.\label{detailedSurveyTable}}
	\begin{footnotesize}
		\begin{tabular}{lrrrrr}
			\toprule[1pt]
			\textbf{Criterion} & $\mathbf{SQL++}$ & $\mathbf{SQL+}$ & $\mathbf{SQL\approx AC}$ & $\mathbf{AC+}$ & $\mathbf{AC++}$ \\
			\midrule[1pt]
			Overall & 0 & 0 & 0 & 3 & 5 \\
			\midrule
			Learning & 0 & 0 & 0 & 2 & 6 \\
			\midrule
			Correct Claims & 0 & 0 & 0 & 1 & 7 \\
			\midrule
			Incorrect Claims & 0 & 0 & 1 & 3 & 4 \\
			\bottomrule[1pt]
		\end{tabular}
	\end{footnotesize}
\end{table}

\begin{table*}
	\center
	\caption{Examples for erroneous claims.\label{erroneousClaimsTable}}
	\begin{footnotesize}
		\begin{tabular}{p{4cm}p{7cm}p{4cm}r}
			\toprule[1pt]
			\textbf{Erroneous Claim} & \textbf{Author Comment} & \textbf{Ground Truth SQL Query} & \makecell[tr]{\textbf{Correct}\\\textbf{Value}} \\
			\midrule[1pt]
			There were only four previous lifetime bans in my database - \textbf{three} were for repeated substance abuse, one was for gambling.~\cite{53814-2} & Yes -- the data was updated on Sept. 22, and the article was originally published on Aug. 28. There's a note at the end of the article, but you're right the article text should also have been updated. & {\scshape SELECT Count(*) FROM nflsuspensions WHERE Games = `indef' AND Category = `substance abuse, repeated offense'} & 4 \\
			\midrule
			Using their campaign fund-raising committees and leadership political action committees separately, the pair have given money to \textbf{64} candidates.~\cite{NYT14-2} & I think you are correct in that it should be 63 candidates in the article, not 64. & {\scshape SELECT CountDistinct(Recipient) FROM eshoopallone} & 63 \\
			\midrule
			\textbf{13\%} of respondents across the globe tell us they are only self-taught.~\cite{Stackoverflow16} & This was a rounding error/typo on our part -- so yes, you're correct. & {\scshape SELECT Percentage(Education) FROM stackoverflow2016 WHERE Education = `i'm self-taught'} & 14 \\
			\bottomrule[1pt]
		\end{tabular}
	\end{footnotesize}
\end{table*}

\begin{figure}
	\begin{tikzpicture}
	\begin{axis}[ybar=0pt, bar width=2pt, width=8.5cm, height=3cm, xlabel={Test Cases}, ylabel={\# Queries}, ylabel near ticks, xlabel near ticks, ymajorgrids, xtick=\empty, legend style={font=\small}, ymode=log]
	\addplot table[x expr=\coordindex, y index=1, col sep=comma] {plots_used/nrPossibleQueries.txt};
	\end{axis}
	\end{tikzpicture}
	\caption{Number of possible query candidates per data set.\label{nrPossibleQueriesFig}}
\end{figure}
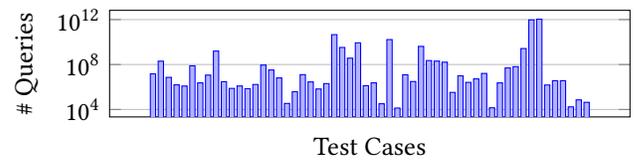

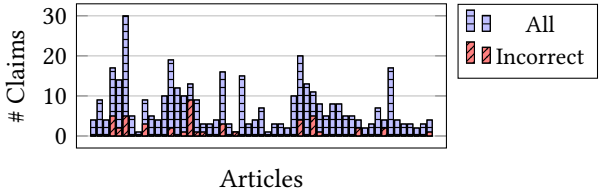
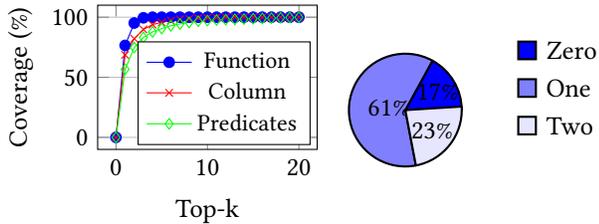
\begin{figure}
	\center
	\subfigure[Distribution of claims over test cases.\label{claimDistributionFig}]{
		\begin{tikzpicture}
		\begin{axis}[ybar=-2pt, bar width=2pt, width=6.5cm, height=3.5cm, xlabel=Articles, ylabel={\# Claims}, ylabel near ticks, xlabel near ticks, ymajorgrids, legend entries={All, Incorrect}, legend pos=outer north east, xtick=\empty, legend style={font=\small}, cycle list name=patternList, enlarge x limits=0.05]
		\addplot table[x expr=\coordindex, y index=3, col sep=comma] {plots_used/annotationTypeNrAllArticle_Correctness.txt};
		\addplot table[x expr=\coordindex, y index=2, col sep=comma] {plots_used/annotationTypeNrAllArticle_Correctness.txt};
		\end{axis}
		\end{tikzpicture}
	}
	\subfigure[Percentage of claims covered per document when considering N most frequent query characteristics.\label{queryCharacteristicsFig}]{
		\begin{tikzpicture}
		\begin{axis}[width=4.5cm, height=3.5cm, xlabel=Top-k, ylabel={Coverage (\%)}, ylabel near ticks, legend entries={Function, Column, Predicates}, legend pos=south east, ymajorgrids, legend style={font=\small}, cycle list name=markerList]
		\addplot table[x index=0, y index=1, col sep=comma] {plots_used/documentTopNCovergeRatio.txt};
		\addplot table[x index=0, y index=2, col sep=comma] {plots_used/documentTopNCovergeRatio.txt};
		\addplot table[x index=0, y index=3, col sep=comma] {plots_used/documentTopNCovergeRatio.txt};
		\end{axis}
		\end{tikzpicture}
	}
	\subfigure[Breakdown of claim queries by number of predicate components.\label{nrPredsFig}]{
		\begin{tikzpicture}
		\pie[radius=0.75, text=legend, color={blue, blue!50, blue!10}]{17/Zero, 61/One, 23/Two}
		\node at (0,-1.5) {};
		\end{tikzpicture}
	}
	\caption{Analysis of test case articles.\label{testCasesFig}}
\end{figure}

\section{Details on Test Cases}
\label{testCasesSub}

We collected 53 publicly available articles summarizing data sets. All test cases will be made available online on the demo website. The most important criterion for our selection was that the article must unambiguously identify a tabular data set it refers to. Under that constraint, we aimed at collecting articles from different sources and authors, treating a variety of topics (covering for instance sports, politics or economy). Our goal was to obtain experimental results that are representative across a variety of domains and writing styles. We used newspaper articles from the New York Times~\cite{NYT}, 538~\cite{538}, Vox~\cite{Vox}, summaries of developer surveys on Stack Overflow~\cite{Stackoverflow}, and Wikipedia~\cite{Wikipedia} articles. The associated data sets range in size from a few kilobytes to around 100 megabytes. Most of them were stored in the .csv format. In a few instance, we removed free text comments, written before or after the actual table data, to obtain valid .csv format. We did however not apply any kind of pre-processing that could have simplified the fact-checking process (e.g., we did not change the original column or value names in the data nor did we change the data structure in any way). 

Table~\ref{erroneousClaimsTable} presents a selection of the erroneous claims we discovered in those test cases. We added comments on likely error causes that we received from the article authors. The primary challenge solved by the AggChecker is the mapping from text claims to SQL queries. This task becomes harder, the more queries can be formed according to our target query structure. Figure~\ref{nrPossibleQueriesFig} shows the number of \textit{Simple Aggregate Queries} (according to the definition in Section~\ref{problemSec}) that can be formed for our test data sets (the three Wikipedia articles reference total of six tables). Evidently, the number of queries is typically enormous, reaching for instance more than a trillion queries for the Stack Overflow Developer Survey 2017~\cite{Stackoverflow17} (this data set features more than 154 table columns). 

To generate ground truth for the claims in our articles, we constructed corresponding SQL queries by hand, after a careful analysis of text and data. We contacted the article authors in case of ambiguities. The AggChecker currently supports claims about \textit{Simple Aggregate Queries} (see Section~\ref{problemSec}). We identified 392 claims that comply with the supported format. The AggChecker is based on the hypothesis that such claims are relatively common in text documents summarizing data. To test that hypothesis, we compared against another type of claim supported by another recently proposed system. The MARGOT system~\cite{LippiT16-2} extracts claims that are backed up not by structured data but by arguments in text. Applying MARGOT to the same set of test cases as the AggChecker, we were able to extract 389 claims. The claim type supported by the AggChecker system is therefore relatively common in comparison. At the same time, such claims are relatively error-prone as shown in Figure~\ref{claimDistributionFig}. We found 12\% of claims to be erroneous and 17 out of 53 test cases contain at least one erroneous claim.




Further, we hypothesize that documents have a primary theme such that queries matching claims in the same document tend to have similar characteristics. This is the motivation behind the expectation maximization approach described in Section~\ref{modelSec}. Figure~\ref{queryCharacteristicsFig} shows the percentage of claim queries per document covered by considering for each query characteristic (i.e., aggregation function, aggregation column, or the column set on which predicates are placed) only the N most frequent instances in the corresponding document (e.g., the three most frequent aggregation functions in the document). Figure~\ref{queryCharacteristicsFig} shows that most queries in a document share the same characteristics (e.g., considering only the top three characteristics covers in average 90.8\% of claims in a document). This confirms our second hypothesis.

Finally, being well aware of limitations of current natural language understanding methods, this project was partially motivated by the hope that claims made in practice tend to be rather simple (i.e., they translate into simple SQL queries). Articles summarizing data sets are often targeted at mainstream users without specific IT background. Avoiding all too complex claims makes articles more readable. Figure~\ref{nrPredsFig} classifies queries by simply counting the number of equality predicates in the corresponding {\scshape WHERE} clause. This is only one out of several ways to measure query complexity. Still, the figure shows that most queries tend to be relatively simple according to that metric.



\begin{figure}
	\center	
	
	\begin{tikzpicture}[mark size=1pt]
	\begin{axis}[width=5cm, height=3cm, xlabel={Top-k Queries}, ylabel={Coverage (\%)}, ylabel near ticks, xlabel near ticks, legend entries={Total, Correct Claims, Incorrect Claims}, legend pos=outer north east, ymajorgrids, legend style={font=\small}, ymin=0, ymax=100, cycle list name=markerList]
	\addplot table[x index=0, y index=1, col sep=comma] {plots_used/topNQueriesRatio_Correctness.txt};
	\addplot table[x index=0, y index=2, col sep=comma] {plots_used/topNQueriesRatio_Correctness.txt};
	\addplot table[x index=0, y index=3, col sep=comma] {plots_used/topNQueriesRatio_Correctness.txt};
	\end{axis}
	\end{tikzpicture}
	
	\caption{Percentage of claims for which correct queries were in the N most likely queries.\label{automatedCoverageFig}}
\end{figure}
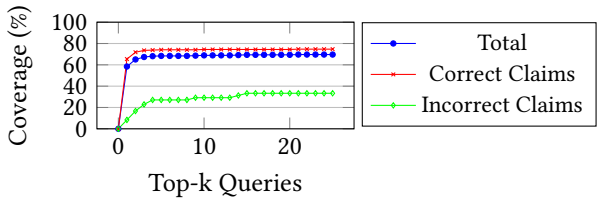

\section{Automated Checking Accuracy}
\label{automatedSub}

We evaluate the AggChecker in fully automated mode. We demonstrate the impact of various design decisions on the text to query translation accuracy. Fully automated fact-checking is not our primary use case as user feedback can help to improve checking accuracy. Still, the first phase of fact-checking (inferring probability distributions over claim queries) is purely automated and can be benchmarked separately. The higher the accuracy of the first phase, the smaller the number of corrective actions required from users.

We use the metric, \textit{Top-k coverage}, which is defined over a set of claims with respect to a positive integer $k$ as follows:
\begin{definition}\textbf{Top-k coverage} is the percentage of claims for which the right query is within the top-k likely query candidates.
\end{definition}

To put the following results into perspective, note that our test cases \textit{allow in average to form $3.76 \times 10^{10}$ query candidates} that comply with our query model. 
Figure~\ref{automatedCoverageFig} shows top-k coverage of claims using fully automated verification.
The most likely query (which is used for tentative fact-checking before user intervention) is in 58.4\% of claims the right one. For 68.4\% of the test claims, the right query is within the top-5 recommendations (each of those queries can be selected with only two clicks total by users).

Besides total coverage, Figure~\ref{automatedCoverageFig} reports top-k coverage for correct and incorrect claims (according to ground truth) separately. Clearly, coverage is higher for correct claims as matching the evaluation result of the query to the text value provides us with strong evidence. Note however that, even if we cannot recognize the right query for an incorrect claim, it will still often be marked as probably incorrect (since no matching query candidate can be found either).

Finally, we validate our design decisions by examining the factors that contribute towards higher accuracy as follows:


\begin{figure}
	\center
	\begin{tikzpicture}
	\begin{axis}[ybar=0, bar width=5pt, width=5cm, height=3.5cm, ylabel={Coverage (\%)}, xticklabels={Top-1, Top5, Top-10}, legend entries={Claim Sentence, +Previous Sentence, +Paragraph Start, +Synonyms, +Headlines}, legend style={font=\small}, legend pos=outer north east, xtick=data, ymajorgrids, enlarge x limits=0.2, ylabel near ticks, cycle list name=patternList]
	\addplot table[x expr=\coordindex, y index=1, col sep=comma] {plots_used/accuracyPerLucene.txt};
	\addplot table[x expr=\coordindex, y index=2, col sep=comma] {plots_used/accuracyPerLucene.txt};
	\addplot table[x expr=\coordindex, y index=3, col sep=comma] {plots_used/accuracyPerLucene.txt};
	\addplot table[x expr=\coordindex, y index=4, col sep=comma] {plots_used/accuracyPerLucene.txt};
	\addplot table[x expr=\coordindex, y index=5, col sep=comma] {plots_used/accuracyPerLucene.txt};
	\end{axis}
	\end{tikzpicture}
	\caption{Top-k coverage as a function of keyword context.\label{keywordsFig}}
\end{figure}
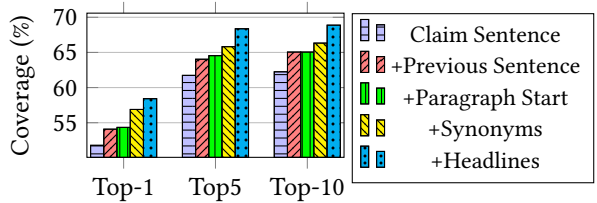

\begin{table}
	\centering
	\caption{Top-k coverage versus probabilistic model.\label{probablisticTable}}
	\begin{small}
		\begin{tabular}{lrrr}
			\toprule[1pt]
			\textbf{Version} & \textbf{Top-1} & \textbf{Top-5} & \textbf{Top-10}\\
			\midrule[1pt]
			Relevance scores $S_c$ & 10.7\% & 31.6\% & 41.1\% \\
			\midrule
			+ Evaluation results $E_c$ & 53.1\% & 64.8\% & 65.8\% \\
			\midrule
			+ Learning priors $\Theta$ & 58.4\% & 68.4\% & 68.9\% \\
			\bottomrule[1pt]
		\end{tabular}
	\end{small}
\end{table}

\textbf{Keyword context.} Figure~\ref{keywordsFig} illustrates the impact of keyword context on text to query translation coverage. Clearly, in particular for determining the most likely query, each keyword source considered by the AggChecker is helpful and improves top-k coverage. 

\textbf{Probabilistic model.}
Table~\ref{probablisticTable} demonstrates the benefits of the probabilistic model. Compared to using keyword-based relevance scores alone (variables $S_c$), we successively improve top-k coverage by first integrating query evaluation results (variables $E_c$) and then document-specific priors (parameters $\Theta$). 


\begin{figure}
	\center
	\begin{tikzpicture}
	\begin{axis}[width=6cm, height=3cm, xlabel={$p_T$}, ylabel={}, ylabel near ticks, xlabel near ticks, legend entries={Recall, Precision, F1 Score}, legend pos=outer north east, ymajorgrids, legend style={font=\small}, cycle list name=markerList]
	\addplot table[x index=0, y index=1, col sep=comma] {plots_used/recallPrecisionTradeOff.txt};
	\addplot table[x index=0, y index=2, col sep=comma] {plots_used/recallPrecisionTradeOff.txt};
	\addplot table[x index=0, y index=3, col sep=comma] {plots_used/recallPrecisionTradeOff.txt};
	\end{axis}
	\end{tikzpicture}
	\caption{Parameter $p_T$ versus recall and precision.\label{pTFig}}
\end{figure}
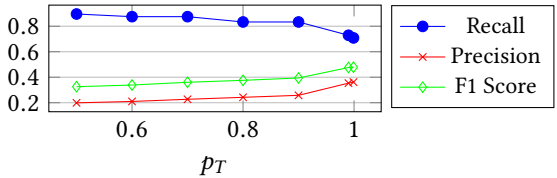

\textbf{Effect of parameter $p_T$.} Parameter $p_T$ is the assumed a-priori probability of encountering correct claims. Figure~\ref{pTFig} shows that we obtain different tradeoffs between recall (i.e., percentage of erroneous claims spotted, based on the most likely query for each claim) and precision (i.e., percentage of claims correctly marked up as wrong) when varying it. Reducing $p_T$ makes the system more ``suspicious'' and increase recall at the cost of precision. We empirically determined $p_T=0.999$ to yield a good tradeoff for our test data set. 

\begin{figure}
	\center
	\begin{tikzpicture}[mark size=1.5pt]
	\begin{groupplot}[width=4.5cm, height=3.5cm, %
	group style={group size = 2 by 1, x descriptions at=edge bottom, ylabels at=edge left, horizontal sep=0.5cm},  xlabel={Time (s)}, ylabel={Coverage (\%)}, xlabel near ticks, ylabel near ticks, ymajorgrids, xmajorgrids, yminorgrids, legend columns=2, legend entries={Top-1, Top-10}, legend to name=tradeoffLegend, cycle list name=markerList]
	\nextgroupplot[title={\# Hits}]
	\addplot table[x index=0, y index=2, col sep=comma] {plots_used/timeAccuracy_HitsPerPage.txt};
	\addplot table[x index=0, y index=3, col sep=comma] {plots_used/timeAccuracy_HitsPerPage.txt};
	\nextgroupplot[title={\# Aggregates}]
	\addplot table[x index=0, y index=2, col sep=comma] {plots_used/timeAccuracy_AggColumn.txt};
	\addplot table[x index=0, y index=3, col sep=comma] {plots_used/timeAccuracy_AggColumn.txt};
	\end{groupplot}
	\end{tikzpicture}
	
	\ref{tradeoffLegend}
	\caption{Top-k coverage versus processing overheads.\label{tradeoffFig}}
\end{figure}
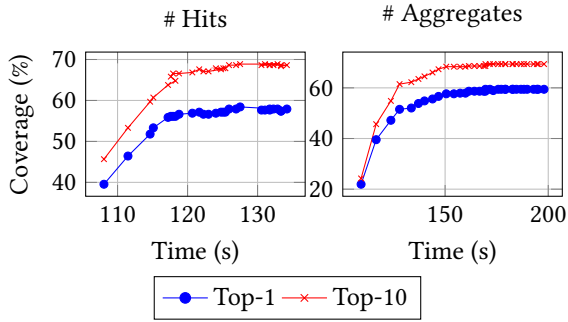

\textbf{Effect of massive processing.}
Figure~\ref{tradeoffFig} shows the benefit of massive processing. We vary the number of hits to collect using Apache Lucene per claim (left) as well as the number of aggregation columns we consider during evaluation (right). Both affect run time (for all test case) as well as average top-k coverage. It turns out that a higher processing time budget results in greater coverage. On the other side, the figure shows that the increase in coverage diminishes as we evaluate more query candidates. Thus as mentioned in Section~\ref{querySec}, evaluating a carefully chosen subset of query candidates is sufficient for achieving high coverage.

\section{Crowd Worker Study}
We conducted an additional larger user study with crowd workers, recruited on Amazon Mechanical Turk (AMT)~\cite{AMTlink}. Our goal was to show that the AggChecker can be used by typical crowd workers (whom we do not expect to have a strong IT background) and without prior training. Furthermore, we compared against yet another baseline that is commonly used by laymen to analyze data: spreadsheets. We uploaded data for all test cases into an online demo of the AggChecker as well as into a public Google Sheet document. 

First, we asked 50 distinct crowd workers to verify numerical claims in a 538 newspaper article~\cite{53814} via the AggChecker. We asked 50 additional workers to verify the same article using Google Sheets. We paid 25 cents per task and did not set any worker eligibility constraints. We compare baselines in terms of recall (i.e., percentage of erroneous claims identified) and precision (i.e., ratio of actually erroneous claims among all claims flagged by crowd worker). We had only 19 respondents for the AggChecker interface and 13 respondents for the Google sheet interface over a 24 hours period. Table~\ref{AMTTable} summarizes the performance results (scope: document). While the performance of crowd workers is only slightly worse compared to the participants of our prior user study when using the AggChecker, crowd workers are unable to identify a single erroneous claim via spreadsheets. 


We doubled the payment and narrowed the scope for verification down to two sentences (taken from another 538 article~\cite{53816-2}). We deliberately selected an article with a very small data set where claims could even be verified by counting entries by hand. All 100 tasks were solved within a 24 hours period. Table~\ref{AMTTable} (scope: paragraph) shows improved results for the spreadsheet, the performance difference to the AggChecker is however enormous.

\begin{table}
	\centering
	\caption{Amazon Mechanical Turk results.\label{AMTTable}}
	\begin{small}
		\begin{tabular}{llrrr}
			\toprule[1pt]
			\textbf{Tool} & \textbf{Scope} & \textbf{Recall} & \textbf{Precision} & \textbf{F1 Score} \\
			\midrule[1pt]
			AggChecker & Document & 56\% & 53\% & 54\% \\
			\midrule
			G-Sheet & & 0\% & 0\% & 0\% \\
			\midrule
			AggChecker & Paragraph & 86\% & 96\% & 91\% \\
			\midrule
			G-Sheet & & 42\% & 95\% & 58\% \\
			\bottomrule[1pt]
		\end{tabular}
	\end{small}
\end{table}

%% file: main.bbl

\begin{thebibliography}{48}


\ifx \showCODEN    \undefined \def \showCODEN     #1{\unskip}     \fi
\ifx \showDOI      \undefined \def \showDOI       #1{#1}\fi
\ifx \showISBNx    \undefined \def \showISBNx     #1{\unskip}     \fi
\ifx \showISBNxiii \undefined \def \showISBNxiii  #1{\unskip}     \fi
\ifx \showISSN     \undefined \def \showISSN      #1{\unskip}     \fi
\ifx \showLCCN     \undefined \def \showLCCN      #1{\unskip}     \fi
\ifx \shownote     \undefined \def \shownote      #1{#1}          \fi
\ifx \showarticletitle \undefined \def \showarticletitle #1{#1}   \fi
\ifx \showURL      \undefined \def \showURL       {\relax}        \fi
\providecommand\bibfield[2]{#2}
\providecommand\bibinfo[2]{#2}
\providecommand\natexlab[1]{#1}
\providecommand\showeprint[2][]{arXiv:#2}

\bibitem[\protect\citeauthoryear{Agrawal, Chaudhuri, and Das}{Agrawal
  et~al\mbox{.}}{2002}]%
        {Agrawal2002}
\bibfield{author}{\bibinfo{person}{Sanjay Agrawal}, \bibinfo{person}{Surajit
  Chaudhuri}, {and} \bibinfo{person}{Gautam Das}.}
  \bibinfo{year}{2002}\natexlab{}.
\newblock \showarticletitle{{DBXplorer: a system for keyword-based search over
  relational databases}}. In \bibinfo{booktitle}{\emph{ICDE}}.
  \bibinfo{pages}{5--16}.
\newblock
\showISBNx{0-7695-1531-2}
\showISSN{1063-6382}
\urldef\tempurl%
\url{https://doi.org/10.1109/ICDE.2002.994693}
\showDOI{\tempurl}


\bibitem[\protect\citeauthoryear{Amazon}{Amazon}{[n. d.]}]%
        {AMTlink}
\bibfield{author}{\bibinfo{person}{Amazon}.} \bibinfo{year}{[n.
  d.]}\natexlab{}.
\newblock \bibinfo{title}{https://www.mturk.com/mturk/welcome}.
\newblock
\newblock


\bibitem[\protect\citeauthoryear{Babakar and Moy}{Babakar and Moy}{2016}]%
        {BabakarM16}
\bibfield{author}{\bibinfo{person}{Mevan Babakar} {and} \bibinfo{person}{Will
  Moy}.} \bibinfo{year}{2016}\natexlab{}.
\newblock \showarticletitle{The state of Automated Factchecking}.
\newblock \bibinfo{journal}{\emph{Full Fact}} (\bibinfo{year}{2016}).
\newblock


\bibitem[\protect\citeauthoryear{Ciampaglia, Shiralkar, Rocha, Bollen, Menczer,
  and Flammini}{Ciampaglia et~al\mbox{.}}{2015}]%
        {CiampagliaS15}
\bibfield{author}{\bibinfo{person}{Giovanni~Luca Ciampaglia},
  \bibinfo{person}{Prashant Shiralkar}, \bibinfo{person}{Luis~Mateus Rocha},
  \bibinfo{person}{Johan Bollen}, \bibinfo{person}{Filippo Menczer}, {and}
  \bibinfo{person}{Alessandro Flammini}.} \bibinfo{year}{2015}\natexlab{}.
\newblock \showarticletitle{Computational fact checking from knowledge
  networks}.
\newblock \bibinfo{journal}{\emph{CoRR}}  \bibinfo{volume}{abs/1501.03471}
  (\bibinfo{year}{2015}).
\newblock


\bibitem[\protect\citeauthoryear{Company}{Company}{2014a}]%
        {NYT14}
\bibfield{author}{\bibinfo{person}{The New York~Times Company}.}
  \bibinfo{year}{2014}\natexlab{a}.
\newblock \bibinfo{title}{Looking for John McCain? Try a Sunday Morning Show}.
\newblock
\newblock
\urldef\tempurl%
\url{https://www.nytimes.com/2014/09/06/upshot/looking-for-john-mccain-try-a-sunday-morning-show.html}
\showURL{%
\tempurl}


\bibitem[\protect\citeauthoryear{Company}{Company}{2014b}]%
        {NYT14-2}
\bibfield{author}{\bibinfo{person}{The New York~Times Company}.}
  \bibinfo{year}{2014}\natexlab{b}.
\newblock \bibinfo{title}{Race in `Waxman' Primary Involves Donating Dollars}.
\newblock
\newblock
\urldef\tempurl%
\url{https://www.nytimes.com/2014/04/24/upshot/race-in-waxman-primary-involves-donating-dollars.html}
\showURL{%
\tempurl}


\bibitem[\protect\citeauthoryear{Company}{Company}{2017}]%
        {NYT}
\bibfield{author}{\bibinfo{person}{The New York~Times Company}.}
  \bibinfo{year}{2017}\natexlab{}.
\newblock \bibinfo{title}{The Upshot}.
\newblock
\newblock
\urldef\tempurl%
\url{https://www.nytimes.com/section/upshot/}
\showURL{%
\tempurl}


\bibitem[\protect\citeauthoryear{contributors}{contributors}{2018}]%
        {Wikipedia}
\bibfield{author}{\bibinfo{person}{Wikipedia contributors}.}
  \bibinfo{year}{2018}\natexlab{}.
\newblock \bibinfo{title}{Wikipedia{,} The Free Encyclopedia}.
\newblock
\newblock
\urldef\tempurl%
\url{https://en.wikipedia.org/}
\showURL{%
\tempurl}


\bibitem[\protect\citeauthoryear{Do and Batzoglou}{Do and Batzoglou}{2008}]%
        {do2008expectation}
\bibfield{author}{\bibinfo{person}{Chuong~B Do} {and} \bibinfo{person}{Serafim
  Batzoglou}.} \bibinfo{year}{2008}\natexlab{}.
\newblock \showarticletitle{What is the expectation maximization algorithm?}
\newblock \bibinfo{journal}{\emph{Nature Biotechnology}} \bibinfo{volume}{26},
  \bibinfo{number}{8} (\bibinfo{year}{2008}), \bibinfo{pages}{897--899}.
\newblock


\bibitem[\protect\citeauthoryear{Fellbaum}{Fellbaum}{1998}]%
        {Fellbaum98}
\bibfield{author}{\bibinfo{person}{C. Fellbaum}.}
  \bibinfo{year}{1998}\natexlab{}.
\newblock \bibinfo{booktitle}{\emph{WordNet: an electronic lexical database}}.
\newblock \bibinfo{publisher}{MIT Press}.
\newblock
\showISBNx{9780262061971}
\showLCCN{97048710}
\urldef\tempurl%
\url{https://books.google.com/books?id=Rehu8OOzMIMC}
\showURL{%
\tempurl}


\bibitem[\protect\citeauthoryear{FiveThirtyEight}{FiveThirtyEight}{2014a}]%
        {53814}
\bibfield{author}{\bibinfo{person}{FiveThirtyEight}.}
  \bibinfo{year}{2014}\natexlab{a}.
\newblock \bibinfo{title}{41 Percent Of Fliers Think You're Rude If You Recline
  Your Seat}.
\newblock
\newblock
\urldef\tempurl%
\url{https://fivethirtyeight.com/features/airplane-etiquette-recline-seat/}
\showURL{%
\tempurl}


\bibitem[\protect\citeauthoryear{FiveThirtyEight}{FiveThirtyEight}{2014b}]%
        {53814-2}
\bibfield{author}{\bibinfo{person}{FiveThirtyEight}.}
  \bibinfo{year}{2014}\natexlab{b}.
\newblock \bibinfo{title}{The NFL's Uneven History Of Punishing Domestic
  Violence}.
\newblock
\newblock
\urldef\tempurl%
\url{https://fivethirtyeight.com/features/nfl-domestic-violence-policy-suspensions/}
\showURL{%
\tempurl}


\bibitem[\protect\citeauthoryear{FiveThirtyEight}{FiveThirtyEight}{2015}]%
        {53815}
\bibfield{author}{\bibinfo{person}{FiveThirtyEight}.}
  \bibinfo{year}{2015}\natexlab{}.
\newblock \bibinfo{title}{Blatter's Reign At FIFA Hasn't Helped Soccer's Poor}.
\newblock
\newblock
\urldef\tempurl%
\url{https://fivethirtyeight.com/features/blatters-reign-at-fifa-hasnt-helped-soccers-poor/}
\showURL{%
\tempurl}


\bibitem[\protect\citeauthoryear{FiveThirtyEight}{FiveThirtyEight}{2016a}]%
        {53816}
\bibfield{author}{\bibinfo{person}{FiveThirtyEight}.}
  \bibinfo{year}{2016}\natexlab{a}.
\newblock \bibinfo{title}{Hip-Hop Is Turning On Donald Trump}.
\newblock
\newblock
\urldef\tempurl%
\url{https://projects.fivethirtyeight.com/clinton-trump-hip-hop-lyrics/}
\showURL{%
\tempurl}


\bibitem[\protect\citeauthoryear{FiveThirtyEight}{FiveThirtyEight}{2016b}]%
        {53816-2}
\bibfield{author}{\bibinfo{person}{FiveThirtyEight}.}
  \bibinfo{year}{2016}\natexlab{b}.
\newblock \bibinfo{title}{Sitting Presidents Give Way More Commencement
  Speeches Than They Used To}.
\newblock
\newblock
\urldef\tempurl%
\url{https://goo.gl/7nuGE9}
\showURL{%
\tempurl}


\bibitem[\protect\citeauthoryear{Foundation}{Foundation}{2017}]%
        {Lucene}
\bibfield{author}{\bibinfo{person}{The Apache~Software Foundation}.}
  \bibinfo{year}{2017}\natexlab{}.
\newblock \bibinfo{title}{Apache Lucene Core}.
\newblock
\newblock
\urldef\tempurl%
\url{https://lucene.apache.org/core/}
\showURL{%
\tempurl}


\bibitem[\protect\citeauthoryear{Group}{Group}{2017}]%
        {Postgres}
\bibfield{author}{\bibinfo{person}{PostgreSQL Global~Development Group}.}
  \bibinfo{year}{2017}\natexlab{}.
\newblock \bibinfo{title}{PostgreSQL}.
\newblock
\newblock
\urldef\tempurl%
\url{https://www.postgresql.org/}
\showURL{%
\tempurl}


\bibitem[\protect\citeauthoryear{Hassan, Arslan, Li, and Tremayne}{Hassan
  et~al\mbox{.}}{2017a}]%
        {HassanA17}
\bibfield{author}{\bibinfo{person}{Naeemul Hassan}, \bibinfo{person}{Fatma
  Arslan}, \bibinfo{person}{Chengkai Li}, {and} \bibinfo{person}{Mark
  Tremayne}.} \bibinfo{year}{2017}\natexlab{a}.
\newblock \showarticletitle{Toward automated fact-checking: detecting
  check-worthy factual claims by ClaimBuster}. In
  \bibinfo{booktitle}{\emph{SIGKDD}}. \bibinfo{pages}{1803--1812}.
\newblock


\bibitem[\protect\citeauthoryear{Hassan, Zhang, Arslan, Caraballo, Jimenez,
  Gawsane, Hasan, Joseph, Kulkarni, Nayak, Sable, Li, and Tremayne}{Hassan
  et~al\mbox{.}}{2017b}]%
        {HassanZ17}
\bibfield{author}{\bibinfo{person}{Naeemul Hassan}, \bibinfo{person}{Gensheng
  Zhang}, \bibinfo{person}{Fatma Arslan}, \bibinfo{person}{Josue Caraballo},
  \bibinfo{person}{Damian Jimenez}, \bibinfo{person}{Siddhant Gawsane},
  \bibinfo{person}{Shohedul Hasan}, \bibinfo{person}{Minumol Joseph},
  \bibinfo{person}{Aaditya Kulkarni}, \bibinfo{person}{Anil~Kumar Nayak},
  \bibinfo{person}{Vikas Sable}, \bibinfo{person}{Chengkai Li}, {and}
  \bibinfo{person}{Mark Tremayne}.} \bibinfo{year}{2017}\natexlab{b}.
\newblock \showarticletitle{ClaimBuster: The first-ever end-to-end
  fact-checking system}.
\newblock \bibinfo{journal}{\emph{PVLDB}} \bibinfo{volume}{10},
  \bibinfo{number}{12} (\bibinfo{year}{2017}).
\newblock


\bibitem[\protect\citeauthoryear{Heilman and Smith}{Heilman and Smith}{2009}]%
        {HeilmanS09}
\bibfield{author}{\bibinfo{person}{Michael Heilman} {and}
  \bibinfo{person}{Noah~A Smith}.} \bibinfo{year}{2009}\natexlab{}.
\newblock \bibinfo{booktitle}{\emph{Question generation via overgenerating
  transformations and ranking}}.
\newblock \bibinfo{type}{{T}echnical {R}eport}. \bibinfo{institution}{CMU,
  Language Technologies Institute}.
\newblock


\bibitem[\protect\citeauthoryear{Heilman and Smith}{Heilman and Smith}{2010}]%
        {HeilmanS10}
\bibfield{author}{\bibinfo{person}{Michael Heilman} {and}
  \bibinfo{person}{Noah~A. Smith}.} \bibinfo{year}{2010}\natexlab{}.
\newblock \showarticletitle{Good question! Statistical ranking for question
  generation}. In \bibinfo{booktitle}{\emph{NACL}}. \bibinfo{pages}{609--617}.
\newblock
\urldef\tempurl%
\url{http://www.aclweb.org/anthology/N10-1086}
\showURL{%
\tempurl}


\bibitem[\protect\citeauthoryear{Inc.}{Inc.}{2017}]%
        {538}
\bibfield{author}{\bibinfo{person}{ESPN Inc.}} \bibinfo{year}{2017}\natexlab{}.
\newblock \bibinfo{title}{FiveThirtyEight}.
\newblock
\newblock
\urldef\tempurl%
\url{http://fivethirtyeight.com/}
\showURL{%
\tempurl}


\bibitem[\protect\citeauthoryear{Karadzhov, Nakov, M{\`{a}}rquez,
  Barr{\'{o}}n{-}Cede{\~{n}}o, and Koychev}{Karadzhov et~al\mbox{.}}{2017}]%
        {KaradzhovN17}
\bibfield{author}{\bibinfo{person}{Georgi Karadzhov}, \bibinfo{person}{Preslav
  Nakov}, \bibinfo{person}{Llu{\'{\i}}s M{\`{a}}rquez},
  \bibinfo{person}{Alberto Barr{\'{o}}n{-}Cede{\~{n}}o}, {and}
  \bibinfo{person}{Ivan Koychev}.} \bibinfo{year}{2017}\natexlab{}.
\newblock \showarticletitle{Fully automated fact checking using external
  sources}.
\newblock \bibinfo{journal}{\emph{CoRR}}  \bibinfo{volume}{abs/1710.00341}
  (\bibinfo{year}{2017}).
\newblock


\bibitem[\protect\citeauthoryear{Li and Jagadish}{Li and Jagadish}{2014a}]%
        {Li2014}
\bibfield{author}{\bibinfo{person}{Fei Li} {and} \bibinfo{person}{HV
  Jagadish}.} \bibinfo{year}{2014}\natexlab{a}.
\newblock \showarticletitle{{NaLIR: an interactive natural language interface
  for querying relational databases}}.
\newblock \bibinfo{journal}{\emph{SIGMOD}} (\bibinfo{year}{2014}),
  \bibinfo{pages}{709--712}.
\newblock
\showISBNx{9781450323765}
\showISSN{07308078}
\urldef\tempurl%
\url{https://doi.org/10.1145/2588555.2594519}
\showDOI{\tempurl}


\bibitem[\protect\citeauthoryear{Li and Jagadish}{Li and Jagadish}{2016}]%
        {Arbor2016}
\bibfield{author}{\bibinfo{person}{Fei Li} {and} \bibinfo{person}{HV
  Jagadish}.} \bibinfo{year}{2016}\natexlab{}.
\newblock \showarticletitle{{Understanding natural language queries over
  relational databases}}.
\newblock \bibinfo{journal}{\emph{SIGMOD Record}} \bibinfo{volume}{45},
  \bibinfo{number}{1} (\bibinfo{year}{2016}), \bibinfo{pages}{6--13}.
\newblock


\bibitem[\protect\citeauthoryear{Li and Jagadish}{Li and Jagadish}{2014b}]%
        {LiJ14}
\bibfield{author}{\bibinfo{person}{Fei Li} {and} \bibinfo{person}{H.~V.
  Jagadish}.} \bibinfo{year}{2014}\natexlab{b}.
\newblock \showarticletitle{Constructing an interactive natural language
  interface for relational databases}.
\newblock \bibinfo{journal}{\emph{{PVLDB}}} \bibinfo{volume}{8},
  \bibinfo{number}{1} (\bibinfo{year}{2014}), \bibinfo{pages}{73--84}.
\newblock
\urldef\tempurl%
\url{http://www.vldb.org/pvldb/vol8/p73-li.pdf}
\showURL{%
\tempurl}


\bibitem[\protect\citeauthoryear{Lippi and Torroni}{Lippi and Torroni}{2016a}]%
        {LippiT16-1}
\bibfield{author}{\bibinfo{person}{Marco Lippi} {and} \bibinfo{person}{Paolo
  Torroni}.} \bibinfo{year}{2016}\natexlab{a}.
\newblock \showarticletitle{Argumentation mining: state of the art and emerging
  trends}.
\newblock \bibinfo{journal}{\emph{{ACM} Trans. Internet Techn.}}
  \bibinfo{volume}{16}, \bibinfo{number}{2} (\bibinfo{year}{2016}),
  \bibinfo{pages}{10:1--10:25}.
\newblock
\urldef\tempurl%
\url{https://doi.org/10.1145/2850417}
\showDOI{\tempurl}


\bibitem[\protect\citeauthoryear{Lippi and Torroni}{Lippi and Torroni}{2016b}]%
        {LippiT16-2}
\bibfield{author}{\bibinfo{person}{Marco Lippi} {and} \bibinfo{person}{Paolo
  Torroni}.} \bibinfo{year}{2016}\natexlab{b}.
\newblock \showarticletitle{{MARGOT:} {A} web server for argumentation mining}.
\newblock \bibinfo{journal}{\emph{Expert Syst. Appl.}}  \bibinfo{volume}{65}
  (\bibinfo{year}{2016}), \bibinfo{pages}{292--303}.
\newblock
\urldef\tempurl%
\url{https://doi.org/10.1016/j.eswa.2016.08.050}
\showDOI{\tempurl}


\bibitem[\protect\citeauthoryear{Manning, Surdeanu, Bauer, Finkel, Bethard, and
  McClosky}{Manning et~al\mbox{.}}{2014}]%
        {ManningS14}
\bibfield{author}{\bibinfo{person}{Christopher~D. Manning},
  \bibinfo{person}{Mihai Surdeanu}, \bibinfo{person}{John Bauer},
  \bibinfo{person}{Jenny~Rose Finkel}, \bibinfo{person}{Steven Bethard}, {and}
  \bibinfo{person}{David McClosky}.} \bibinfo{year}{2014}\natexlab{}.
\newblock \showarticletitle{The Stanford CoreNLP Natural Language Processing
  Toolkit}. In \bibinfo{booktitle}{\emph{ACL}}. \bibinfo{pages}{55--60}.
\newblock


\bibitem[\protect\citeauthoryear{Media}{Media}{2017}]%
        {Vox}
\bibfield{author}{\bibinfo{person}{Vox Media}.}
  \bibinfo{year}{2017}\natexlab{}.
\newblock \bibinfo{title}{Vox}.
\newblock
\newblock
\urldef\tempurl%
\url{https://www.vox.com/}
\showURL{%
\tempurl}


\bibitem[\protect\citeauthoryear{Miller}{Miller}{1995}]%
        {Miller95}
\bibfield{author}{\bibinfo{person}{George~A. Miller}.}
  \bibinfo{year}{1995}\natexlab{}.
\newblock \showarticletitle{WordNet: a lexical database for English}.
\newblock \bibinfo{journal}{\emph{Commun. {ACM}}} \bibinfo{volume}{38},
  \bibinfo{number}{11} (\bibinfo{year}{1995}), \bibinfo{pages}{39--41}.
\newblock


\bibitem[\protect\citeauthoryear{Nakashole and Mitchell}{Nakashole and
  Mitchell}{2014}]%
        {NakasholeM14}
\bibfield{author}{\bibinfo{person}{Ndapandula Nakashole} {and}
  \bibinfo{person}{Tom~M. Mitchell}.} \bibinfo{year}{2014}\natexlab{}.
\newblock \showarticletitle{Language-aware truth assessment of fact
  candidates}. In \bibinfo{booktitle}{\emph{ACL}}. \bibinfo{pages}{1009--1019}.
\newblock


\bibitem[\protect\citeauthoryear{Peldszus and Stede}{Peldszus and
  Stede}{2013}]%
        {PeldszusS13}
\bibfield{author}{\bibinfo{person}{Andreas Peldszus} {and}
  \bibinfo{person}{Manfred Stede}.} \bibinfo{year}{2013}\natexlab{}.
\newblock \showarticletitle{From argument diagrams to argumentation mining in
  texts: a survey}.
\newblock \bibinfo{journal}{\emph{{IJCINI}}} \bibinfo{volume}{7},
  \bibinfo{number}{1} (\bibinfo{year}{2013}), \bibinfo{pages}{1--31}.
\newblock
\urldef\tempurl%
\url{https://doi.org/10.4018/jcini.2013010101}
\showDOI{\tempurl}


\bibitem[\protect\citeauthoryear{Saha, Floratou, Sankaranarayanan, Minhas,
  Mittal, and Ozcan}{Saha et~al\mbox{.}}{2016}]%
        {Saha2016}
\bibfield{author}{\bibinfo{person}{Diptikalyan Saha}, \bibinfo{person}{Avrilia
  Floratou}, \bibinfo{person}{Karthik Sankaranarayanan},
  \bibinfo{person}{Umar~Farooq Minhas}, \bibinfo{person}{Ashish~R Mittal},
  {and} \bibinfo{person}{Fatma Ozcan}.} \bibinfo{year}{2016}\natexlab{}.
\newblock \showarticletitle{{ATHENA: An ontology-driven system for natural
  language querying over relational data stores}}.
\newblock \bibinfo{journal}{\emph{VLDB}} \bibinfo{volume}{9},
  \bibinfo{number}{12} (\bibinfo{year}{2016}), \bibinfo{pages}{1209--1220}.
\newblock


\bibitem[\protect\citeauthoryear{Shi and Weninger}{Shi and Weninger}{2015}]%
        {ShiW15}
\bibfield{author}{\bibinfo{person}{Baoxu Shi} {and} \bibinfo{person}{Tim
  Weninger}.} \bibinfo{year}{2015}\natexlab{}.
\newblock \showarticletitle{Fact checking in large knowledge graphs - a
  discriminative predicate path mining approach}.
\newblock \bibinfo{journal}{\emph{CoRR}}  \bibinfo{volume}{abs/1510.05911}
  (\bibinfo{year}{2015}).
\newblock


\bibitem[\protect\citeauthoryear{Shi and Weninger}{Shi and Weninger}{2016}]%
        {ShiW16}
\bibfield{author}{\bibinfo{person}{Baoxu Shi} {and} \bibinfo{person}{Tim
  Weninger}.} \bibinfo{year}{2016}\natexlab{}.
\newblock \showarticletitle{Fact checking in heterogeneous information
  networks}. In \bibinfo{booktitle}{\emph{WWW}}. \bibinfo{pages}{101--102}.
\newblock


\bibitem[\protect\citeauthoryear{Stack~Exchange}{Stack~Exchange}{2015}]%
        {Stackoverflow15}
\bibfield{author}{\bibinfo{person}{Inc. Stack~Exchange}.}
  \bibinfo{year}{2015}\natexlab{}.
\newblock \bibinfo{title}{2015 Developer Survey}.
\newblock
\newblock
\urldef\tempurl%
\url{https://insights.stackoverflow.com/survey/2015/}
\showURL{%
\tempurl}


\bibitem[\protect\citeauthoryear{Stack~Exchange}{Stack~Exchange}{2016}]%
        {Stackoverflow16}
\bibfield{author}{\bibinfo{person}{Inc. Stack~Exchange}.}
  \bibinfo{year}{2016}\natexlab{}.
\newblock \bibinfo{title}{Developer Survey Results 2016}.
\newblock
\newblock
\urldef\tempurl%
\url{https://insights.stackoverflow.com/survey/2016/}
\showURL{%
\tempurl}


\bibitem[\protect\citeauthoryear{Stack~Exchange}{Stack~Exchange}{2017a}]%
        {Stackoverflow17}
\bibfield{author}{\bibinfo{person}{Inc. Stack~Exchange}.}
  \bibinfo{year}{2017}\natexlab{a}.
\newblock \bibinfo{title}{Developer Survey Results 2017}.
\newblock
\newblock
\urldef\tempurl%
\url{https://insights.stackoverflow.com/survey/2017/}
\showURL{%
\tempurl}


\bibitem[\protect\citeauthoryear{Stack~Exchange}{Stack~Exchange}{2017b}]%
        {Stackoverflow}
\bibfield{author}{\bibinfo{person}{Inc. Stack~Exchange}.}
  \bibinfo{year}{2017}\natexlab{b}.
\newblock \bibinfo{title}{Stack Overflow Insights}.
\newblock
\newblock
\urldef\tempurl%
\url{https://insights.stackoverflow.com/survey/}
\showURL{%
\tempurl}


\bibitem[\protect\citeauthoryear{Thorne and Vlachos}{Thorne and
  Vlachos}{2017}]%
        {ThorneV17}
\bibfield{author}{\bibinfo{person}{James Thorne} {and} \bibinfo{person}{Andreas
  Vlachos}.} \bibinfo{year}{2017}\natexlab{}.
\newblock \showarticletitle{An extensible framework for verification of
  numerical claims}. In \bibinfo{booktitle}{\emph{EACL}}.
  \bibinfo{pages}{37--40}.
\newblock


\bibitem[\protect\citeauthoryear{Vlachos and Riedel}{Vlachos and
  Riedel}{2014}]%
        {VlachosR14}
\bibfield{author}{\bibinfo{person}{Andreas Vlachos} {and}
  \bibinfo{person}{Sebastian Riedel}.} \bibinfo{year}{2014}\natexlab{}.
\newblock \showarticletitle{Fact Checking: Task definition and dataset
  construction}.
\newblock \bibinfo{journal}{\emph{ACL Workshop on Language Technologies and
  Computational Social Science}}, \bibinfo{pages}{18--22}.
\newblock


\bibitem[\protect\citeauthoryear{Vlachos and Riedel}{Vlachos and
  Riedel}{2015}]%
        {VlachosR15}
\bibfield{author}{\bibinfo{person}{Andreas Vlachos} {and}
  \bibinfo{person}{Sebastian Riedel}.} \bibinfo{year}{2015}\natexlab{}.
\newblock \showarticletitle{Identification and verification of simple claims
  about statistical properties}. In \bibinfo{booktitle}{\emph{EMNLP}}.
  \bibinfo{pages}{2596--2601}.
\newblock


\bibitem[\protect\citeauthoryear{Walenz and Yang}{Walenz and Yang}{2016}]%
        {WalenzY16}
\bibfield{author}{\bibinfo{person}{Brett Walenz} {and} \bibinfo{person}{Jun
  Yang}.} \bibinfo{year}{2016}\natexlab{}.
\newblock \showarticletitle{Perturbation analysis of database queries}.
\newblock \bibinfo{journal}{\emph{PVLDB}} \bibinfo{volume}{9},
  \bibinfo{number}{14} (\bibinfo{year}{2016}), \bibinfo{pages}{1635--1646}.
\newblock
\urldef\tempurl%
\url{http://www.vldb.org/pvldb/vol9/p1635-walenz.pdf}
\showURL{%
\tempurl}


\bibitem[\protect\citeauthoryear{Wang}{Wang}{2017}]%
        {Wang17}
\bibfield{author}{\bibinfo{person}{William~Yang Wang}.}
  \bibinfo{year}{2017}\natexlab{}.
\newblock \showarticletitle{"Liar, Liar Pants on Fire": {A} new benchmark
  dataset for fake news detection}. In \bibinfo{booktitle}{\emph{ACL}}.
  \bibinfo{pages}{422--426}.
\newblock


\bibitem[\protect\citeauthoryear{Wu, Agarwal, Li, Yang, and Yu}{Wu
  et~al\mbox{.}}{2014a}]%
        {WuA14}
\bibfield{author}{\bibinfo{person}{You Wu}, \bibinfo{person}{Pankaj~K.
  Agarwal}, \bibinfo{person}{Chengkai Li}, \bibinfo{person}{Jun Yang}, {and}
  \bibinfo{person}{Cong Yu}.} \bibinfo{year}{2014}\natexlab{a}.
\newblock \showarticletitle{Toward computational fact-checking}.
\newblock \bibinfo{journal}{\emph{PVLDB}} \bibinfo{volume}{7},
  \bibinfo{number}{7} (\bibinfo{year}{2014}), \bibinfo{pages}{589--600}.
\newblock


\bibitem[\protect\citeauthoryear{Wu, Agarwal, Li, Yang, and Yu}{Wu
  et~al\mbox{.}}{2017}]%
        {WuA17}
\bibfield{author}{\bibinfo{person}{You Wu}, \bibinfo{person}{Pankaj~K.
  Agarwal}, \bibinfo{person}{Chengkai Li}, \bibinfo{person}{Jun Yang}, {and}
  \bibinfo{person}{Cong Yu}.} \bibinfo{year}{2017}\natexlab{}.
\newblock \showarticletitle{Computational fact checking through query
  perturbations}.
\newblock \bibinfo{journal}{\emph{TODS}} \bibinfo{volume}{42},
  \bibinfo{number}{1} (\bibinfo{year}{2017}), \bibinfo{pages}{4:1--4:41}.
\newblock
\urldef\tempurl%
\url{https://doi.org/10.1145/2996453}
\showDOI{\tempurl}


\bibitem[\protect\citeauthoryear{Wu, Walenz, Li, Shim, Sonmez, Agarwal, Li,
  Yang, and Yu}{Wu et~al\mbox{.}}{2014b}]%
        {WuW14}
\bibfield{author}{\bibinfo{person}{You Wu}, \bibinfo{person}{Brett Walenz},
  \bibinfo{person}{Peggy Li}, \bibinfo{person}{Andrew Shim},
  \bibinfo{person}{Emre Sonmez}, \bibinfo{person}{Pankaj~K. Agarwal},
  \bibinfo{person}{Chengkai Li}, \bibinfo{person}{Jun Yang}, {and}
  \bibinfo{person}{Cong Yu}.} \bibinfo{year}{2014}\natexlab{b}.
\newblock \showarticletitle{iCheck: computationally combating "lies, d-ned
  lies, and statistics"}. In \bibinfo{booktitle}{\emph{SIGMOD}}.
  \bibinfo{pages}{1063--1066}.
\newblock


\end{thebibliography}
